\documentclass[12pt]{article}
\makeatletter\if@twocolumn\PassOptionsToPackage{switch}{lineno}\else\fi\makeatother

\usepackage{amsfonts,amssymb,amsbsy,latexsym,amsmath,tabulary,graphicx,times,caption,fancyhdr,xfrac}
\usepackage[utf8]{inputenc}
\usepackage[nolists,nomarkers]{endfloat}

\usepackage{url,multirow,morefloats,floatflt,cancel,tfrupee}
\makeatletter

\AtBeginDocument{\@ifpackageloaded{textcomp}{}{\usepackage{textcomp}}}
\makeatother
\usepackage{colortbl}
\usepackage{xcolor}
\usepackage{pifont}
\usepackage[nointegrals]{wasysym}
\makeatletter

\def\mcWidth#1{\csname TY@F#1\endcsname+\tabcolsep}

\def\mathLarge#1{\mbox{\Large $#1$}}

\def\cAlignHack{\rightskip\@flushglue\leftskip\@flushglue\parindent\z@\parfillskip\z@skip}
\def\rAlignHack{\rightskip\z@skip\leftskip\@flushglue \parindent\z@\parfillskip\z@skip}

\@ifundefined{etal}{}{}

\usepackage{ifxetex}
\ifxetex\else\if@twocolumn\@ifpackageloaded{stfloats}{}{\usepackage{dblfloatfix}}\fi\fi

\AtBeginDocument{
\expandafter\ifx\csname eqalign\endcsname\relax
\def\eqalign#1{\null\vcenter{\def\\{\cr}\openup\jot\m@th
  \ialign{\strut$\displaystyle{##}$\hfil&$\displaystyle{{}##}$\hfil
      \crcr#1\crcr}}\,}
\fi
}

\AtBeginDocument{%
  \@ifpackageloaded{endfloat}%
   {\renewcommand\efloat@iwrite[1]{\immediate\expandafter\protected@write\csname efloat@post#1\endcsname{}}}{\newif\ifefloat@tables}%
}%

\def\BreakURLText#1{\@tfor\brk@tempa:=#1\do{\brk@tempa\hskip0pt}}
\let\lt=<
\let\gt=>
\def\processVert{\ifmmode|\else\textbar\fi}

\@ifundefined{subparagraph}{
\def\subparagraph{\@startsection{paragraph}{5}{2\parindent}{0ex plus 0.1ex minus 0.1ex}%
{0ex}{\normalfont\small\itshape}}%
}{}

\newcommand\role[1]{\unskip}
\newcommand\aucollab[1]{\unskip}
  
\@ifundefined{tsGraphicsScaleX}{\gdef\tsGraphicsScaleX{1}}{}
\@ifundefined{tsGraphicsScaleY}{\gdef\tsGraphicsScaleY{.9}}{}
\def\checkGraphicsWidth{\ifdim\Gin@nat@width>\linewidth
	\tsGraphicsScaleX\linewidth\else\Gin@nat@width\fi}

\def\checkGraphicsHeight{\ifdim\Gin@nat@height>.9\textheight
	\tsGraphicsScaleY\textheight\else\Gin@nat@height\fi}

\def\fixFloatSize#1{}
\let\ts@includegraphics\includegraphics

\def\inlinegraphic[#1]#2{{\edef\@tempa{#1}\edef\baseline@shift{\ifx\@tempa\@empty0\else#1\fi}\edef\tempZ{\the\numexpr(\numexpr(\baseline@shift*\f@size/100))}\protect\raisebox{\tempZ pt}{\ts@includegraphics{#2}}}}

\AtBeginDocument{\def\includegraphics{\@ifnextchar[{\ts@includegraphics}{\ts@includegraphics[width=\checkGraphicsWidth,height=\checkGraphicsHeight,keepaspectratio]}}}

\DeclareMathAlphabet{\mathpzc}{OT1}{pzc}{m}{it}

\def\URL#1#2{\@ifundefined{href}{#2}{\href{#1}{#2}}}

\def\UrlOrds{\do\*\do\-\do\~\do\'\do\"\do\-}%
\g@addto@macro{\UrlBreaks}{\UrlOrds}

\edef\fntEncoding{\f@encoding}

\makeatother

\newif\ifmultipleabstract\multipleabstractfalse%
\makeatletter

\def\wileyIndent{1pt}
\usepackage[paperheight=11.69in,paperwidth=8.27in,margin=3cm,headsep=.5cm,top=2.5cm]{geometry}

\renewenvironment{abstract}
{\vspace*{-1pc}\trivlist\item[]\leftskip\wileyIndent\hrulefill\par\vskip4pt\noindent\textbf{\abstractname}\mbox{\null}\\}{\par\noindent\hrulefill\endtrivlist}

\def\author#1{\gdef\@author{\hskip-\dimexpr(\tabcolsep)\hskip\wileyIndent\parbox{\dimexpr\textwidth-\wileyIndent}{\raggedright\itshape#1}}}

\def\title#1{\gdef\@title{\raggedright\bfseries\ifx\@articleType\@empty\else\@articleType\\\fi#1}}

\let\@articleType\@empty \def\articletype#1{\gdef\@articleType{{\normalfont\itshape#1}}}

\fancypagestyle{headings}{\fancyhf{}\fancyhead[C]{\RunningHead}\fancyhead[R]{\thepage}}

 \def\audegree#1{}

\usepackage[noindentafter]{titlesec}
\def\NormalBaseline{\def\baselinestretch{1.1}}

\titleformat{\section}[hang]{\NormalBaseline\filright\boldmath\bfseries}
{\thesection.}
{6pt}
{}
[]
\titleformat{\subsection}[hang]{\NormalBaseline\filright\boldmath\bfseries}
{\thesubsection.}
{6pt}
{}
[]
\titleformat{\subsubsection}[hang]{\NormalBaseline\filright\itshape}
{\thesubsubsection.}
{6pt}
{}
[]
\titleformat{\paragraph}[runin]{\NormalBaseline\itshape}
{}
{6pt}
{}
[:]
\titleformat{\subparagraph}[runin]{\NormalBaseline\itshape}
{}
{6pt}
{}
[:]

\titlespacing{\section}{0pt}{1.5\baselineskip}{.2\baselineskip}  
\titlespacing{\subsection}{0pt}{1.5\baselineskip}{.2\baselineskip}  
\titlespacing{\subsubsection}{0pt}{1.5\baselineskip}{.2\baselineskip}  
\titlespacing{\paragraph}{0pt}{.5\baselineskip}{10pt}  
\titlespacing{\subparagraph}{0pt}{.5\baselineskip}{10pt}

\date{}

\makeatother
  
\usepackage[T1]{fontenc}
\makeatother

\usepackage[
autocite    = superscript,
style = chem-angew,
url = false,
doi = false,
isbn = false,
eprint = false,
maxbibnames = 99,
]{biblatex}

\DeclareFieldFormat{pages}{\mkfirstpage{#1}}

\usepackage{jabbrv}

\def\thanksspace{{\phantom{\textsuperscript{\thefootnote}}}}
\usepackage{float}

\newcommand{\texttildeapprox}{{\fontfamily{pcr}\selectfont\texttildelow}}

\begin{document}

\title{Soft mechanical metamaterials with transformable topology protected by stress caching}
\author{Jason~Christopher Jolly\textsuperscript{1}, Binjie~Jin\textsuperscript{2}, Lishuai~Jin\textsuperscript{1}, YoungJoo~Lee\textsuperscript{1}, Tao~Xie\textsuperscript{2}, Stefano~Gonella\textsuperscript{3}, Kai~Sun\textsuperscript{4}, Xiaoming~Mao\textsuperscript{4}\thanks{Corresponding author.}\thanksspace \space and Shu~Yang\textsuperscript{1}\footnotemark[1]\thanksspace ~\\[-3pt]\normalsize\normalfont ~\\
\textsuperscript{1}{Department of Materials Science and Engineering\unskip, University of Pennsylvania\unskip, 3231 Walnut Street\unskip, Philadelphia\unskip, 19104\unskip, Pennsylvania\unskip, USA}~\\
\textsuperscript{2}{State Key Laboratory of Chemical Engineering, Department of Chemical and Biological Engineering\unskip, Zhejiang University\unskip, 38 Zhe Da Road\unskip, Hangzhou\unskip, 310027\unskip, Zhejiang\unskip, China}~\\
\textsuperscript{3}{Department of Civil, Environmental, and Geo- Engineering\unskip, University of Minnesota\unskip, 500 Pillsbury Drive S.E.\unskip, Minneapolis\unskip, 55455\unskip, Minnesota\unskip, USA}~\\
\textsuperscript{4}{Department of Physics\unskip, University of Michigan\unskip, 450 Church St.\unskip, Ann Arbor\unskip, 48109\unskip, Michigan\unskip, USA\newline Corresponding E-mail: maox@umich.edu (Xiaoming~Mao), shuyang@seas.upenn.edu (Shu~Yang)}}

\def\RunningHead{}\def\RunningAuthor{Christopher Jolly \MakeLowercase{\textit{et al.}} }

\maketitle

\begin{abstract}
Maxwell lattice metamaterials possess a rich phase space with distinct topological states featuring mechanically polarized edge behaviors and strongly asymmetric acoustic responses. Until now, demonstrations of non-trivial topological behaviors from Maxwell lattices have been limited to either monoliths with locked configurations or reconfigurable mechanical linkages. This work introduces a transformable topological mechanical metamaterial (TTMM) made from a shape memory polymer and based on a generalized kagome lattice. It is capable of reversibly exploring topologically distinct phases of the non-trivial phase space via a kinematic strategy that converts sparse mechanical inputs at free edge pairs into a biaxial, global transformation that switches its topological state. Thanks to the shape memory effect, all configurations are stable even in the absence of confinement or a continuous mechanical input. Topologically-protected mechanical behaviors, while robust against structural (with broken hinges) or conformational defects (up to $\sim$55\% mis-rotations), are shown to be vulnerable to the adverse effects of stored elastic energy from prior transformations (up to a $\sim$70\% reduction in edge stiffness ratios, depending on hinge width). Interestingly, we show that shape memory polymer's intrinsic phase transitions that modulate chain mobility can effectively shield a dynamic metamaterial's topological response (with a 100\% recovery) from its own kinematic stress history, an effect we refer to as "stress caching".\def\keywordstitle{Keywords}

\end{abstract}

\section{Introduction}
Mechanical metamaterials have extrinsic properties that transcend conventional material behavior including variable Poisson's ratios\unskip~\autocite{Farzaneh2022SequentialRatios,Janbaz2020StrainMetamaterials}, vanishing shear moduli\unskip~\autocite{Kadic2012OnMetamaterials}, 
high stiffness at low densities\unskip~\autocite{Zheng2014UltralightMetamaterials} , tunable stiffness\unskip~\autocite{Florijn2014ProgrammableMetamaterials,Hwang2018TunableStructures,Montgomery2021Magneto-MechanicalBandgaps}, tunable acoustic behavior\unskip~\autocite{Montgomery2021Magneto-MechanicalBandgaps}, 
chirality\unskip~\autocite{Frenzel2017Three-dimensionalTwist} and predetermined or energy dissipative failure\unskip~\autocite{Paulose2015SelectiveMetamaterials,Frenzel2016TailoredAbsorbers}. Such exotic mechanical behaviors facilitate their potential application such as mechanical cloaks, vibration isolators, energy dissipators and switchable acoustic diodes\unskip~\autocite{Paulose2015SelectiveMetamaterials,Frenzel2016TailoredAbsorbers,Zhang2021TailoredIsolation,Shan2015MultistableEnergy,Bilal2017a}. Recently, a subset known as topological mechanical metamaterials have gained attention as the mechanical analogs of electronic topological insulators, in which mesoscopic features in the bulk and their topology in reciprocal space can inform and preserve mechanical behaviors along surfaces and edges or a localized defect\unskip~\autocite{Susstrunk2016ClassificationMetamaterials}. 
Therefore, their physical properties are robust against disorder and wear, which are ubiquitous in any real-world realization of such systems.

Freely-jointed elastic frames known as Maxwell lattices exhibit such topological mechanical phenomena. They have exactly the amount of bonds (\textit{n}\ensuremath{_{B}}) required to balance out the degrees of freedom ($n.d$) of $n$ joints in a $d$ dimensional space i.e., they are critically coordinated with a coordination number, ${z=\frac{2n_B}{n}=2d}$ and are therefore on the brink of instability\unskip~\autocite{Mao2018}. Whereas coordination number (z) and Maxwell's criterion\unskip~\autocite{Maxwell1864L.Frames} 
determine the onset of rigidity at the mean field level, Calladine's counting rule\unskip~\autocite{Pellegrino1986} rigorously accounts for states of self-stress and floppy modes or mechanisms. The latter can be thought of as mechanical charge equivalents of particle-hole pairs, and correspond to displacements of the lattice sites that do not produce strain in the bonds (i.e., floppy modes) and tensile or compressive stresses in bonds that can exist in the absence of equilibrating forces applied at the sites (i.e., self-stress), respectively\unskip~\autocite{Mao2018}. For instance, in an infinite regular kagome lattice (a type of Maxwell lattice) under periodic boundary conditions, sample traversing lines of bonds can accommodate states of self stress and floppy modes\unskip~\autocite{Lubensky2015}. A finite regular kagome lattice sample excised from an infinite sheet has floppy modes courtesy of the cleaved bonds, and these can either reside in the bulk or migrate to the surfaces upon twisting the lattice units to distort the straight lines of bonds\unskip~\autocite{Sun2012}. 
Kane and Lubensky\unskip~\autocite{Kane2013} show how these floppy modes can migrate away from one edge and localize at an opposite edge, resulting in a polarized elastic response that is informed by the topology of the lattice's phonon bands. The localization of these modes is captured by a topological polarization vector, ${{\vec{R}_T}}$ that points along the primitive vectors of the lattice, to the edge that is deemed floppy\unskip~\autocite{Kane2013}. Theoretical predictions and experimental observations of such polarized topological phenomena at both zero and finite frequencies, have been made in two- (2D) and three-dimensional (3D) frames of elastic members\unskip~\autocite{Paulose2015SelectiveMetamaterials,Bilal2017,Baardink2017,Paulose2015TopologicalMetamaterials,Stenull2016TopologicalDimensions,Chen2014NonlinearInsulator,Rocklin2016}, origami and kirigami structures\unskip~\autocite{Chen2016TopologicalKirigami}, and gear assemblies\unskip~\autocite{Meeussen2016GearedStability}. 

Reconfigurability endows metamaterials with tunability and pluripotency, whereby structures capable of altering their topology and modulating protected properties can take advantage of a wider operational phase space than their `fixed-design' counterparts\unskip~\autocite{Haghpanah2016MultistableMaterials}.  
Rocklin et al.\unskip~\autocite{Rocklin2017} proposed a generalized deformed kagome lattice (GDKL) capable of switching between multiple auxetic and topologically polarized states via a global soft transformation. Prior experimental realizations of metamaterials with transformable topology have focused on geometric topological changes, leveraging mechanical compaction of macroscale lattice units\unskip~\autocite{Montgomery2021Magneto-MechanicalBandgaps,Coulais2018Multi-stepMetamaterials} and capillary forces to `zip' microstructures\unskip~\autocite{Li2021Liquid-inducedMicrostructures}, in order
to alter nodal connectivity and modulate static and dynamic mechanical responses. However, investigations into systems with polar elasticity protected by the topology of their phonon bands, have thus far been limited to monoliths with fixed configurations from within the available phase space\unskip~\autocite{Ma2018,Pishvar2020}, or to manually assembled systems wherein reconfigurability is achieved via macroscale mechanical linkages\unskip~\autocite{Rocklin2017}, thereby intrinsically limiting their potential for scalability and miniaturization.

In this work, we show for the first time how topological polarization can be modulated experimentally in a monolithic transformable metamaterial.  We overcome an intrinsic material limitation associated with the conventional elasticity of elastomers, thermoplastics or thermosets, which would otherwise preclude reconfigurability, by fabricating our transformable topological mechanical metamaterial (TTMM) out of a shape memory polymer. A top-down photolithography process allows for the miniaturization of these structures for potential downstream implementation as devices, while offering a high level of control over the smallest, critical hinge features that are essential to realize topologically polarized states. Prescribing precise kinematic transformations in a high degree-of-freedom geometry is  non-trivial. Rather than adopting a brute-force approach of prescribing local unit rotations, our solution, informed by iterative finite element modeling (FEM), leverages an intrinsic Guest-Hutchinson mode\unskip~\autocite{Guest2003OnStructures} 
to achieve this complex biaxial global metamorphosis via only a pair of uniaxial inputs applied at the edges. With handed scissor mechanisms at the sample edges, we suppress kinematic indeterminacy from undesirable buckling modes that arise courtesy of the emergence of sample-traversing aligned bonds at phase boundaries. Here, the intrinsic shape memory effect affords two crucial functionalities: (i) the stability of every conformation without the need for a continuous mechanical input and (ii) the ability to temporarily lock away (i.e., cache) stored elastic energy from a prior kinematic transformation. We validate the importance of the latter through experimental demonstrations of the muted polar elastic response in reference samples made from commercial elastomers which are incapable of caching stress, revealing that unmanaged stored stresses can significantly attenuate topological polarization (up to 70\%). This work provides a blueprint for future realizations of monolithic mechanical metamaterials capable of switching between topological states through the use of functional soft materials. We underscore the robustness of topological edge modes against defects and disorder while providing strategies to circumvent their vulnerability to stored elastic energy.

\section{Results}

\bgroup
\fixFloatSize{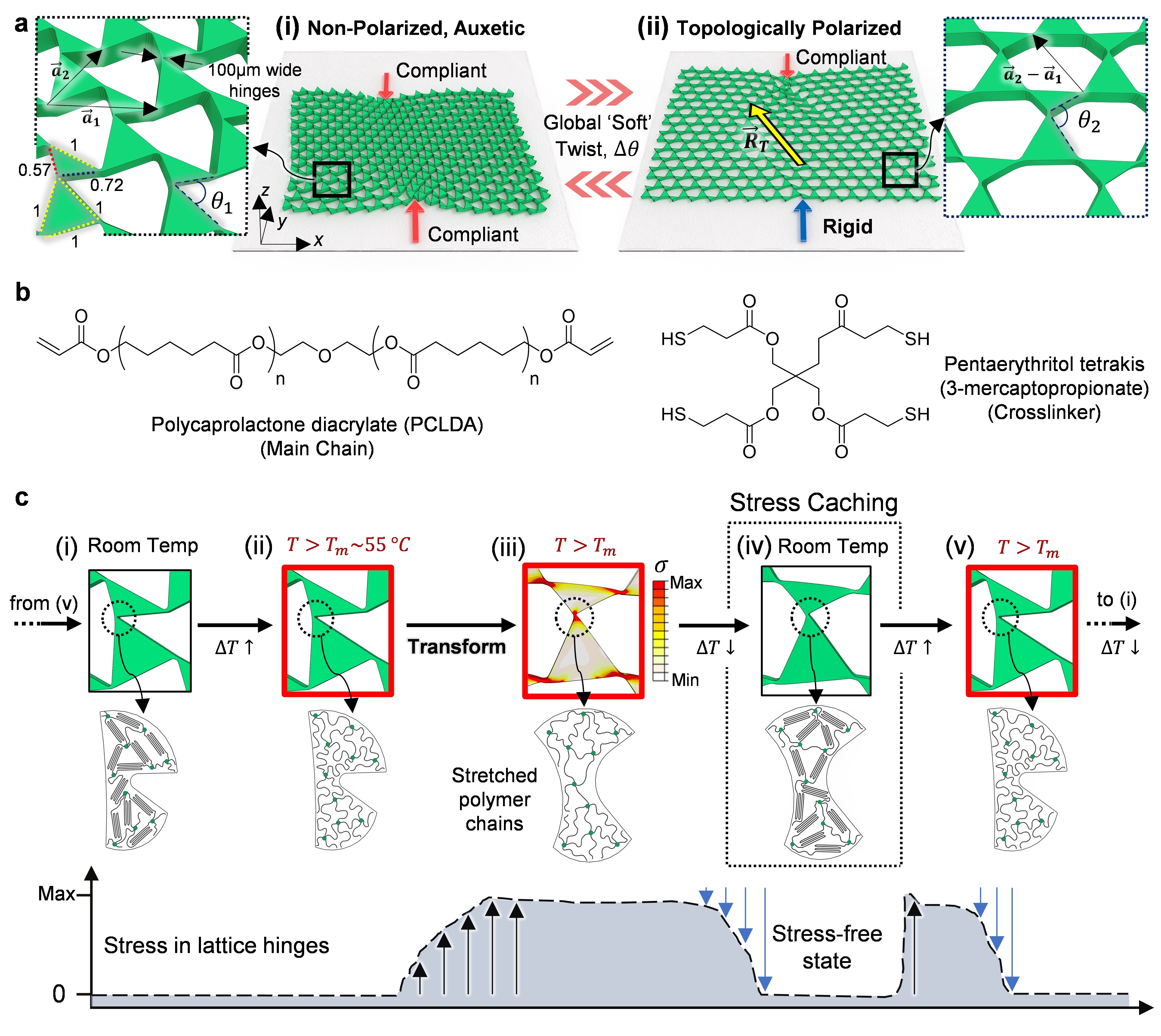}
\begin{figure}[!htbp]
\centering \makeatletter\IfFileExists{Figures/Figure_1.jpg}{\includegraphics{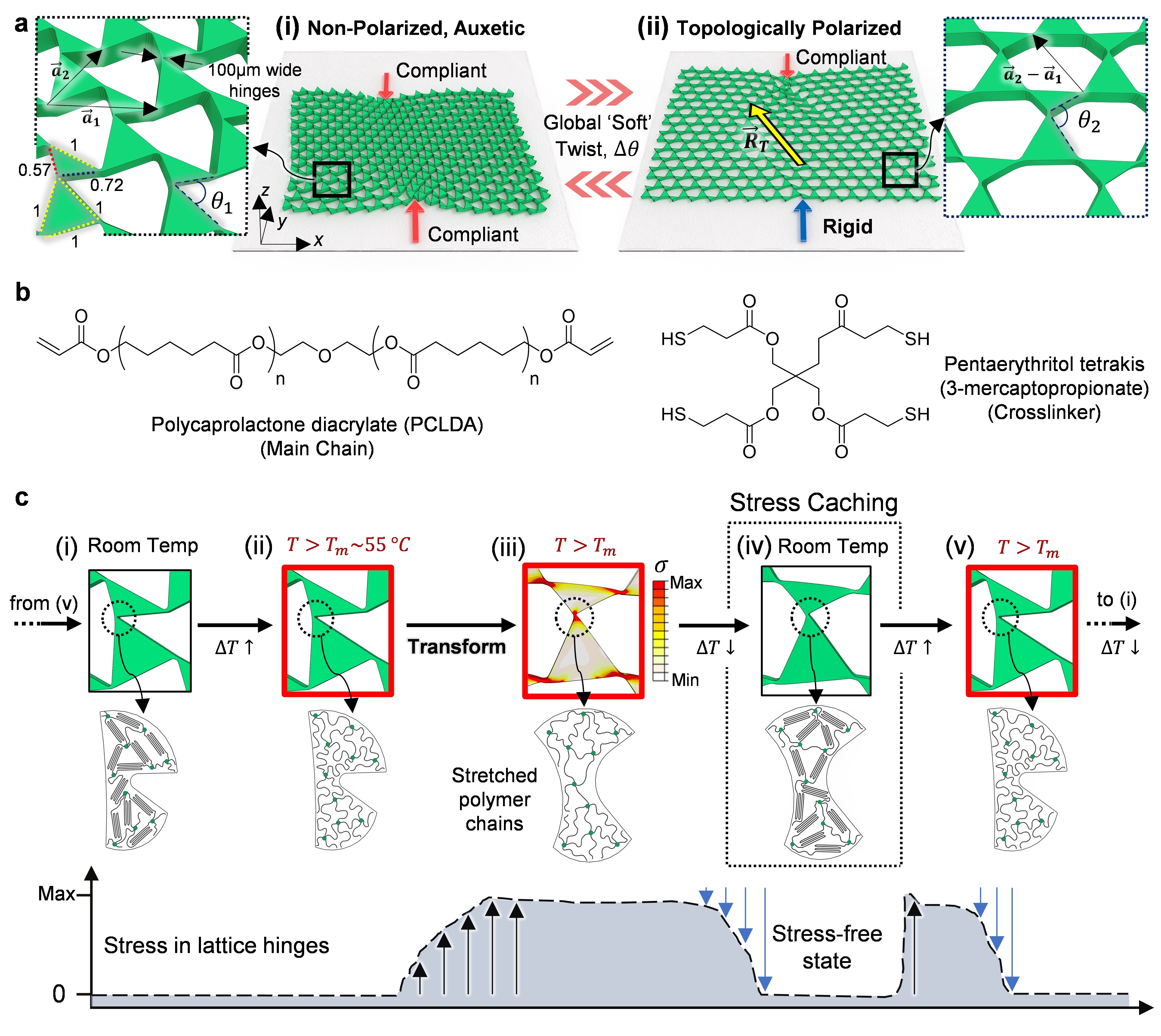}}{}
\makeatother 
\caption{{A TTMM made from a shape memory polymer, PCLDA-SMP, capable of stress caching. a) A `deformed’ kagome lattice TTMM can be reversibly transformed between (i) an auxetic phase with uniformly distributed floppy modes and (ii) a polarized phase with a non-zero topological polarization vector, $\vec{R}_T$ and a polar mechanical response, by uniformly twisting its triangular sub-units. b) Chemical structures of the constituents of the polycaprolactone diacrylate (PCLDA)-based shape memory polymer (PCLDA-SMP). c) Illustrations of the intrinsic shape memory effect: i-iii) Melting crystalline domains in the polymer network facilitates kinematic transformations that store minimal elastic strain energy. iv) The reformation of these domains upon cooling stabilizes the TTMM in each phase even after the applied external input or constraint is removed. The temporary zero-stress state referred to as stress-caching, permits the lattice to circumvent detrimental effects on its topologically polarized edge response, from stresses stored during its transformation. v,i) Remelting the crystalline domains restores polymer mobility and thereby unlocks these cached stresses which, in the absence of external confinement, induces shape recovery to the initial configuration}}
\label{Figure_1}
\end{figure}
\egroup

\subsection{Transformable metamaterials made from functional, soft materials}
The TTMM reported here is based on a 2D GDKL variant that has been shown\unskip~\autocite{Rocklin2017} to have multiple auxetic and topologically polarized conformations depending on the angle $\theta$ subtended by its triangular units. The lattice is constructed by tessellating corner-sharing scalene and equilateral triangles with side length ratios (0.57, 1, 0.72) and (1, 1, 1) respectively, and a maximum side length of 2.25 mm, along primitive vectors  $\vec{a_1}$ and $\vec{a_2} $ (see \textbf{Figure~\ref{Figure_1}}a).  While theoretical treatments\unskip~\autocite{Rocklin2017,Lubensky2015,Kane2013} 
assumed `ideal' or `free' hinges with zero bending stiffness, the triangular repeating units in this work are connected via slender hinge ligaments (100 \ensuremath{\mu}m wide), corresponding to a hinge-width to side-length slenderness ratio of 0.044. This value embodies an optimal compromise between extrinsic compliance (so as to preserve the vestiges of theoretical topological behavior in the continuum elasticity limit), and sample robustness (to survive fabrication, handling and experimentation). 

As described in \cite{Rocklin2017}, the chosen GDKL passes through three topological phase transitions as the angle subtended by the triangular sub-units, $\theta$, increases from $0^\circ$ to $199^\circ$. It exhibits four distinct phases - two auxetic and two polarized. At the extremities of the lattice's rotational phase space, it exists in dilatory, maximally auxetic states, i.e., when $0^\circ<\theta<76.554^\circ$ and $120.211^\circ<\theta<199^\circ$ wherein $\vec{R}_T=0$ and floppy modes are uniformly distributed between all edges (Figure~\ref{Figure_1}a (i)). Soft twisting the lattice into either of its two polarized phases where $78.994^\circ<\theta<94.022^\circ$ and $94.022^\circ<\theta<120^\circ$, sees its surface floppy modes migrate away from a particular edge in a given edge pair (`left-right' edges and `top-bottom' edges, respectively) and localize at its opposite counterpart with $\vec{R}_T=\vec{a_2}-\vec{a_1}$ or $\vec{R}_T=\vec{a_2}$, respectively (Figure~\ref{Figure_1}a (ii)). At the topological phase transitions at $\theta=\theta_{a_1-a_2}=78.994^\circ$, $\theta_{a_1}=94.022^\circ$ and $\theta_{a_2}=120^\circ$ surface floppy modes are temporarily converted into bulk modes due to states of self stress associated with sample spanning straight lines of bonds i.e., aligned triangle edges. These `transient' bulk floppy modes pose a challenge to prescribing a wholly determinate kinematic transformation, which will be addressed later. It must be noted that precise `critical' angles in the ideal lattice become slightly `blurred' in the limit of finite hinge widths and in the presence of disorder. Also, since the two auxetic and two polarized phases that exist on either side of the phase boundaries at $\theta_{a_1-a_2}$ and $\theta_{a_2}$ in the GDKL's rotational phase space are very similar in nature, the discussion henceforth focuses on reversibly transforming the lattice between its auxetic and polarized states on either side of $\theta_{a_1-a_2}$ and characterizing its behavior in each phase. However, we note that our experimental strategy and findings are not limited to this subset phase space nor this particular GDKL geometry and can be adapted to suit other Maxwell lattices with a Guest mode.

In any `real' lattice with hinges of finite width, where free rotations of triangular units are replaced by flexure of hinge ligaments, the associated elasticity could potentially affect its behavior following a kinematic phase transformation. Notably, restorative stresses generated in hinges with finite bending stiffness, $K=EC\kappa t^{3}$, where $E$ is the Young's modulus, $C$ is a geometric constant, $\kappa$ is its curvature and $t$ is the hinge width, would require a continuous mechanical input at the sample boundaries to stabilize each topologically-distinct conformation\unskip~\autocite{Wu2015DirectingUnits}. To free our TTMM from this encumbrance, we imbue it with shape memory. Figure~\ref{Figure_1}b shows the chemical structures of the constituents of our polycaprolactone diacrylate (PCLDA)-based shape memory polymer (PCLDA-SMP). The lattices are made via a multi-step `thick' photolithography process (see detailed process in Figure S1, Supporting Information), optimized to minimize fabrication defects such as broken hinges and missing units and achieve uniform hinge width distributions within $\pm$7 $\mu$m of the targeted 100 $\mu$m value.  Figure~\ref{Figure_1}c shows the general sequence of operations required to transform the TTMM, where the inset shows the morphology of the polymer network in the hinges at each step. At room temperature, the lattice is rendered rigid and elastic by the crystalline domains in the polymer network, and has a Young's modulus, $E\sim132$ MPa (see Figure S2, Supporting Information). The lattice is then heated to above the melting temperature of PCLDA-SMP, $T_m=55^\circ C $ (see Figure S3, Supporting Information), where the material behaves like a conventional soft elastomer with a Young's modulus, $E\sim2.1$ MPa.  In this state, the auxetic (or polarized) lattice is malleable and can be easily transformed into its adjacent topological phase via a stretching (compression) dominated global kinematic transformation that stores a mixture of tensile (compressive) and bending stresses primarily in the hinges as evidenced in the FEM stress heat map shown in Figure~\ref{Figure_1}c (iii). At the molecular level (see the inset schematic), polymer chains are mobile and align with the local stress fields. It must be noted that the temperature modulation of the material's elastic modulus by more than two orders of magnitude exacts a lower elastic energy cost for the lattice transformation, with minimal stress being generated compared to the case if the same transformation was performed at room temperature where the material is stiffer or if a conventional rigid material was used. The transformation of the lattice is completed and locked by cooling down to room temperature without altering its boundary conditions. PCLDA is recrystallized within the polymer network, restoring intrinsic stiffness to the material and `freezing' locally deformed polymer chain conformations to lock the prescribed lattice structure via a shape memory effect. In this state, polymer mobility is effectively zero with relaxation times far exceeding practical experimental time scales\unskip~\autocite{Cho2019}. This allows for locking away or \textit{caching} any stress generated during the transformation of the lattice. Upon reheating the sample to above its T\ensuremath{_{m}}, polymer mobility is restored and the previously cached elastic energy drives its recovery to the initial, unstressed or entropically-favored state. The evolution of localized stresses in the lattice hinges during a typical phase transformation cycle of the lattice, is captured in the inset in Figure~\ref{Figure_1}c.

This stress caching ability is confirmed experimentally via dynamic mechanical testing (Figure S4a, Supporting Information) wherein a PCLDA-SMP sample is heated and subjected to a tensile strain of 105\% (the largest local strain generated in a transformed lattice, from FEM). The measured stress increases with the applied strain up to a maximum equilibrium value and then falls to zero as the sample is cooled down to room temperature. Upon reheating above T\ensuremath{_{m}}, the measured stress recovers sharply to its original value. To confirm that this intrinsic ability is non-trivial, we perform an identical test on a reference sample made from a commercial elastomer, Elite Double 32 (ED-32; Zhermack SpA), which does not exhibit a shape memory effect (Figure S4b, Supporting Information). Not surprisingly, only a mild reduction in stress is measured in the strained ED-32 sample during temperature cycling, which is typical for rubber elasticity. Shape fixity ratio, $R_f = \mathLarge{\sfrac{\epsilon_{f}}{\epsilon_{l}}}\times100\%$, quantifies a material's shape memory abilities, where $\epsilon_{l}$ and $\epsilon_{f}$ are the respective strains before and after removing the applied load\unskip~\autocite{Zhao2016}. Shape memory and stress caching operations are repeated multiple times without any degradation of the PCLDA-SMP material and with 99.8\% shape fixity over at least three cycles, as verified by cyclic dynamic mechanical analysis under stress control (Figure S5, Supporting Information).

\bgroup
\fixFloatSize{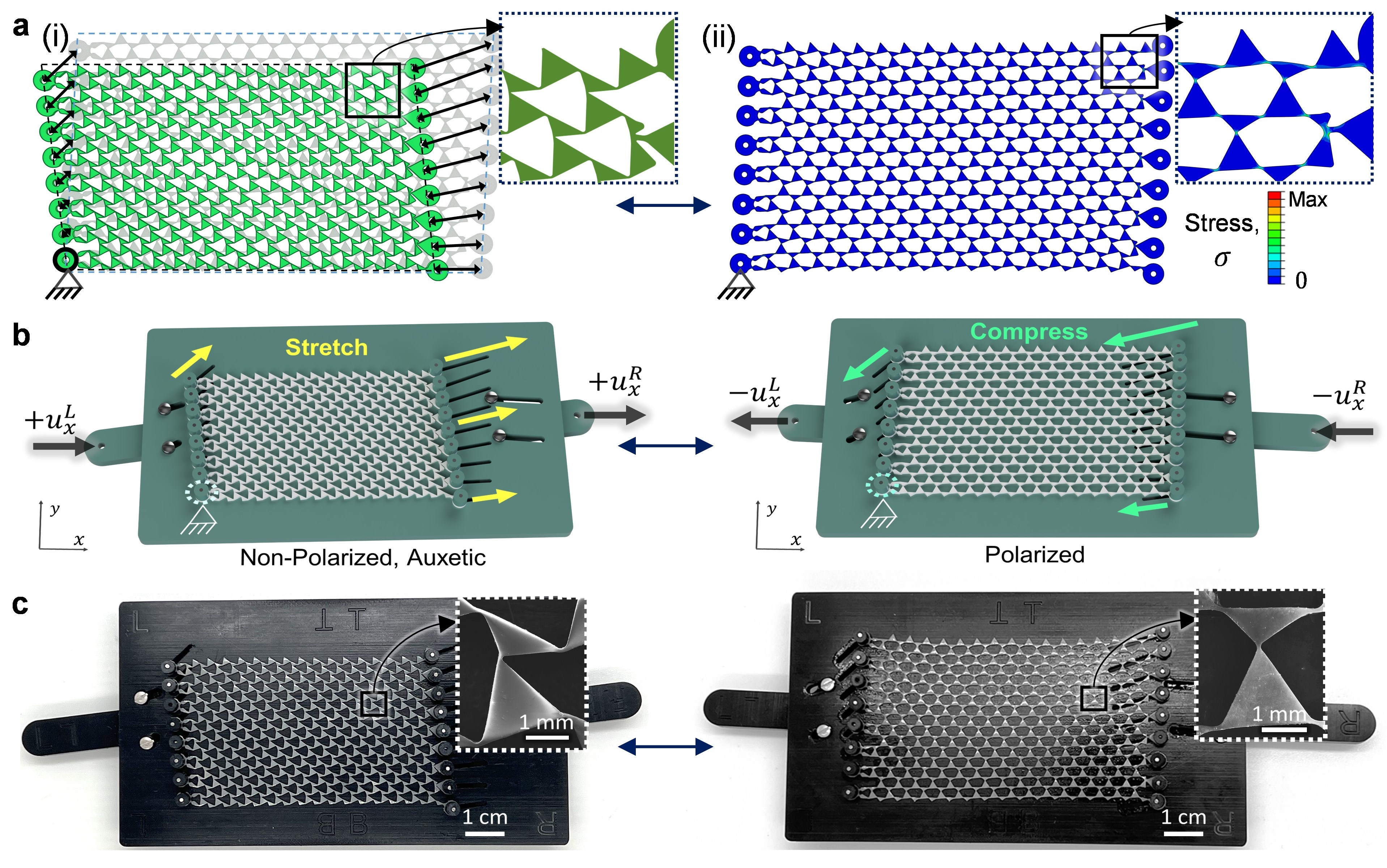}
\begin{figure}[!htbp]
\centering \makeatletter\IfFileExists{Figures/Figure_2.jpg}{\includegraphics{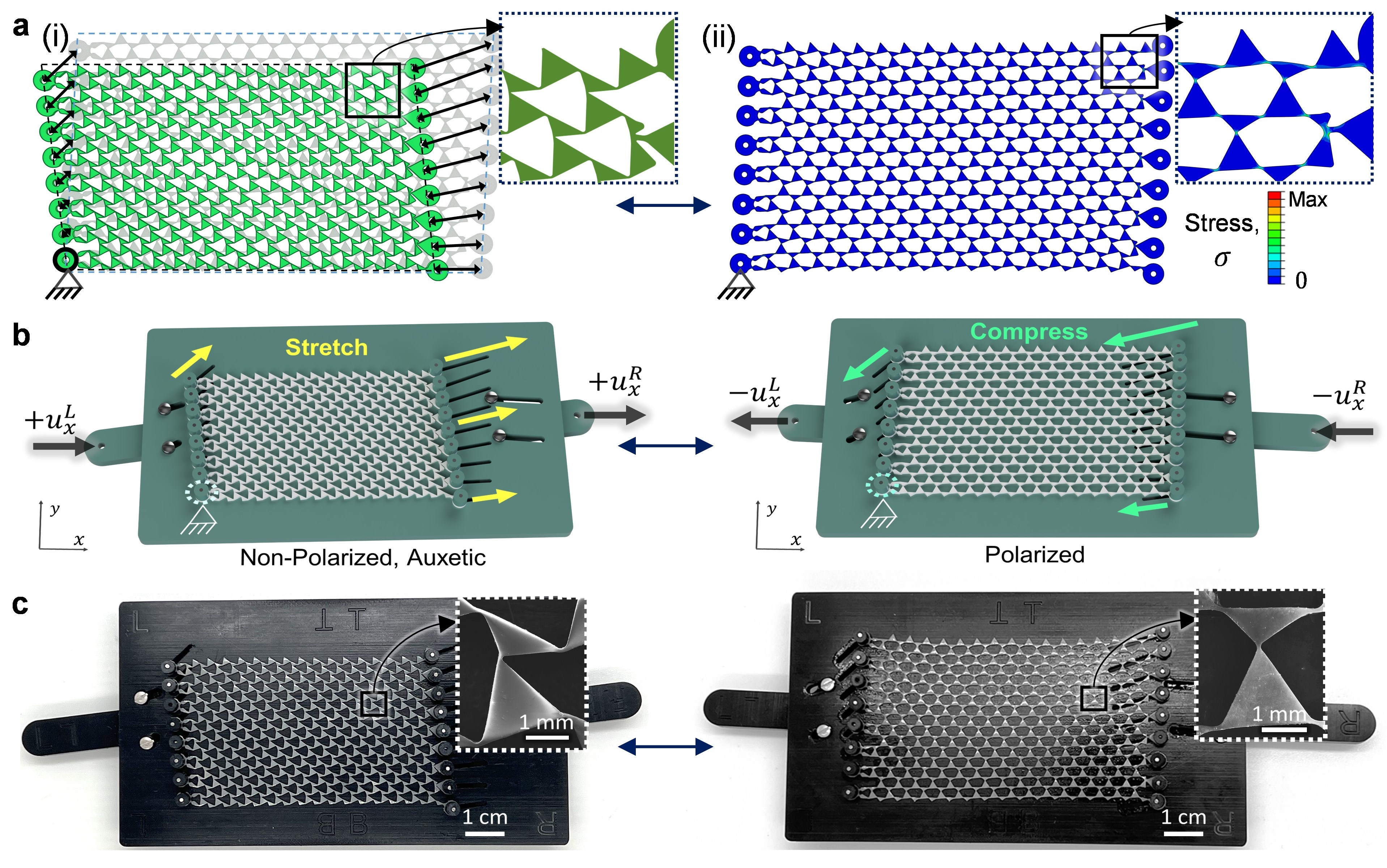}}{}
\makeatother 
\caption{{A kinematic strategy to achieve the biaxial transformation of a high degree-of-freedom lattice, with simple uniaxial mechanical inputs at sample edges. a) (i) Mapping the edges of an initial lattice configuration to those of a superimposed target conformation, determines the requisite vector displacements of the input edges for a reversible transformation of the entire lattice via its Guest mode. (ii) FEM results verifying the effectiveness of the proposed kinematic strategy in polarizing an initially-auxetic lattice. The inset stress color map reveals that stresses generated during the transformation are localized in the hinge ligaments. b) Illustration of a unique jig design, wherein simple uniaxial inputs can be cascaded into synchronously applied displacements of edge loops along laser cut slots, to prescribe a biaxial topological phase transformation of the lattice. c) Optical images of a PCLDA-SMP TTMM lattice sample actuated via pin-in-slot joints in a laser cut PTFE kinematic jig. Inset: SEM images of a single mm-scale repeating unit of the 1 mm thick and cm-scale, free-standing PCLDA-SMP TTMM lattice monolith with 100 $\mu$m wide hinge ligaments.}}
\label{Figure_2}
\end{figure}
\egroup

\subsection{Prescribing kinematic phase transformations}
In order to experimentally prescribe a reversible transformation of the lattice, we need to overcome several challenges: (i) dictating uniform local deformations of all lattice units i.e., triggering the inherent Guest-Hutchinson mode, only via the edges, (ii) the tendency of the lattice's surface floppy modes to localize a naively applied deformation within a finite distance of the input edges, (iii) designing a physical jig that can accommodate both the dilation-dominant behavior of the auxetic phases and the shear-dominant behavior of the polarized phases, (iv) the significant miniaturization of the jig and all its components for the small lattice sizes, and (v) mounting the lattice in the jig for mechanical testing without impeding or biasing test results. Informed by iterative FEM, we choose the left and right (L-R) edge pair as the control edges (the input) while the top and bottom (T-B) edges are left unmodified to measure the lattice's topological edge behavior (the output). The modifications to the input edges involve the addition of loops connected via right- and left-handed scissor mechanisms to the repeating units along the L-R edges, respectively, to eliminate buckling indeterminacy at certain phase boundaries. As shown in \textbf{Figure~\ref{Figure_2}}a (i), individual vector displacements are applied to each edge loop, which collectively map the lattice edges to a target configuration. The rest of the lattice then follows the collective synchronized displacement of its edges, courtesy the Guest-Hutchinson mode. The success of this strategy in prescribing a reversible transformation between an auxetic and a polarized state is verified via FEM before proceeding with experiments (Figure~\ref{Figure_2} a (ii) and Supplementary Movie S1, Supporting Information). While the directions and magnitudes of the individual components in a vector map are dependent on the initial choice of a common, zero-displacement point (in this case, the bottom left corner edge loop), we show in FEM that a successful and reversible transformation can be realized regardless of this initial choice (Figure S6 and Movie S2, Supporting Information). Essentially, for a given final configuration, the behavior of a polarized or auxetic lattice is independent of how it is transformed.

Experimentally, we realize the transformation using a polytetrafluoroethylene (PTFE) jig with laser cut slots that represent the vector displacements of the edge loops (see Figure S7, Supporting Information, and Experimental Section). The PCLDA-SMP lattice is mounted in the jig via `pin-in-slot' joints (Figure~\ref{Figure_2} b and c). Importantly, the jig has `slotted guides' that move in their own vector mapped slots at either edge and work to cascade two simple uniaxial inputs ($u_x^{L}$ and $u_x^{R}$) into synchronized displacements of all edge loops along their individual slots. This in turn triggers the inherent Guest mode of the lattice, wherein the bulk of the lattice emulates the prescribed deformation at the edges, resulting in a homogeneous biaxial kinematic transformation of the lattice. Movie S3, Supporting Information, shows a complete kinematic cycle of the TTMM.

\bgroup
\fixFloatSize{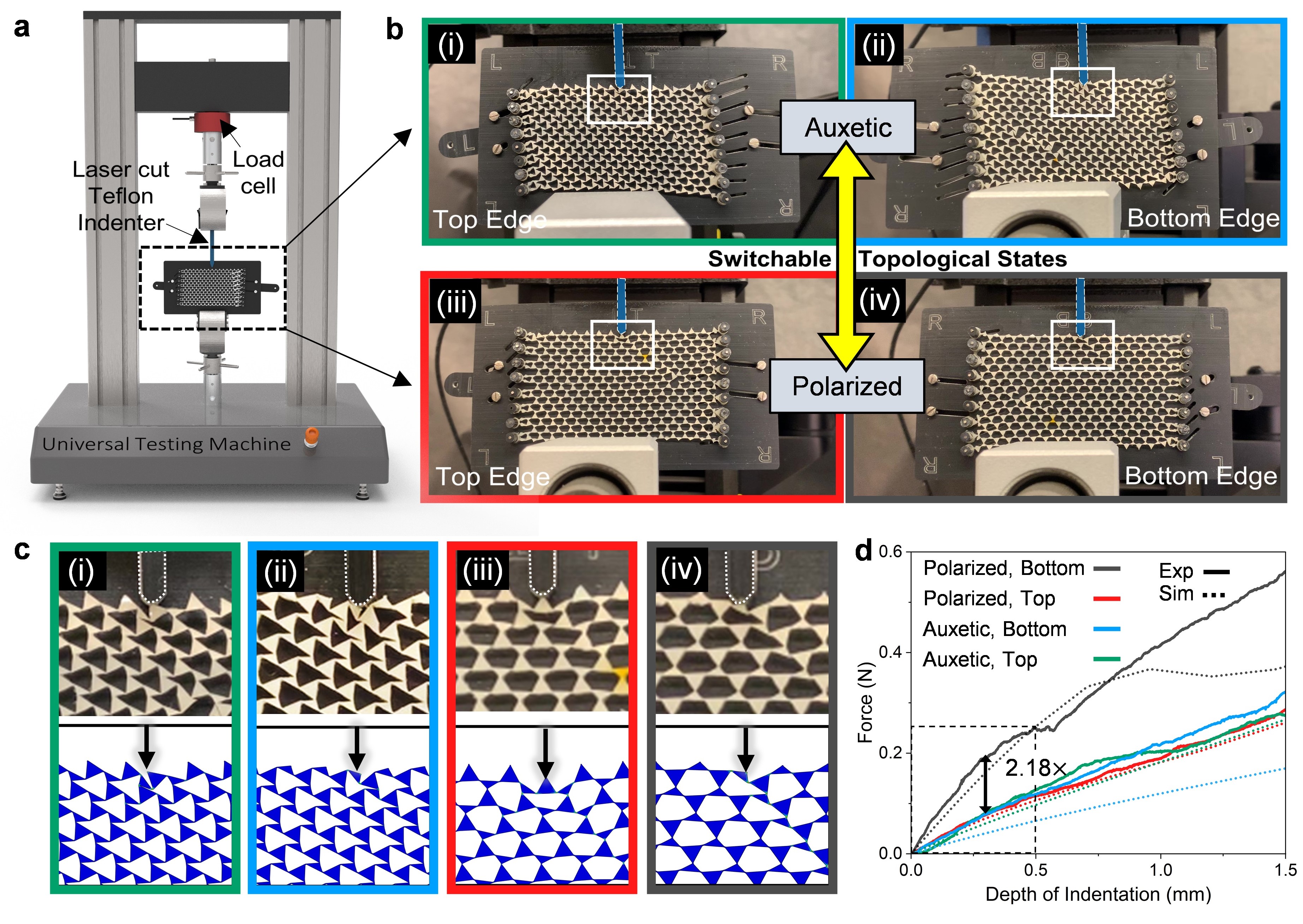}
\begin{figure}[!htbp]
\centering \makeatletter\IfFileExists{Figures/Figure_3.jpg}{\includegraphics{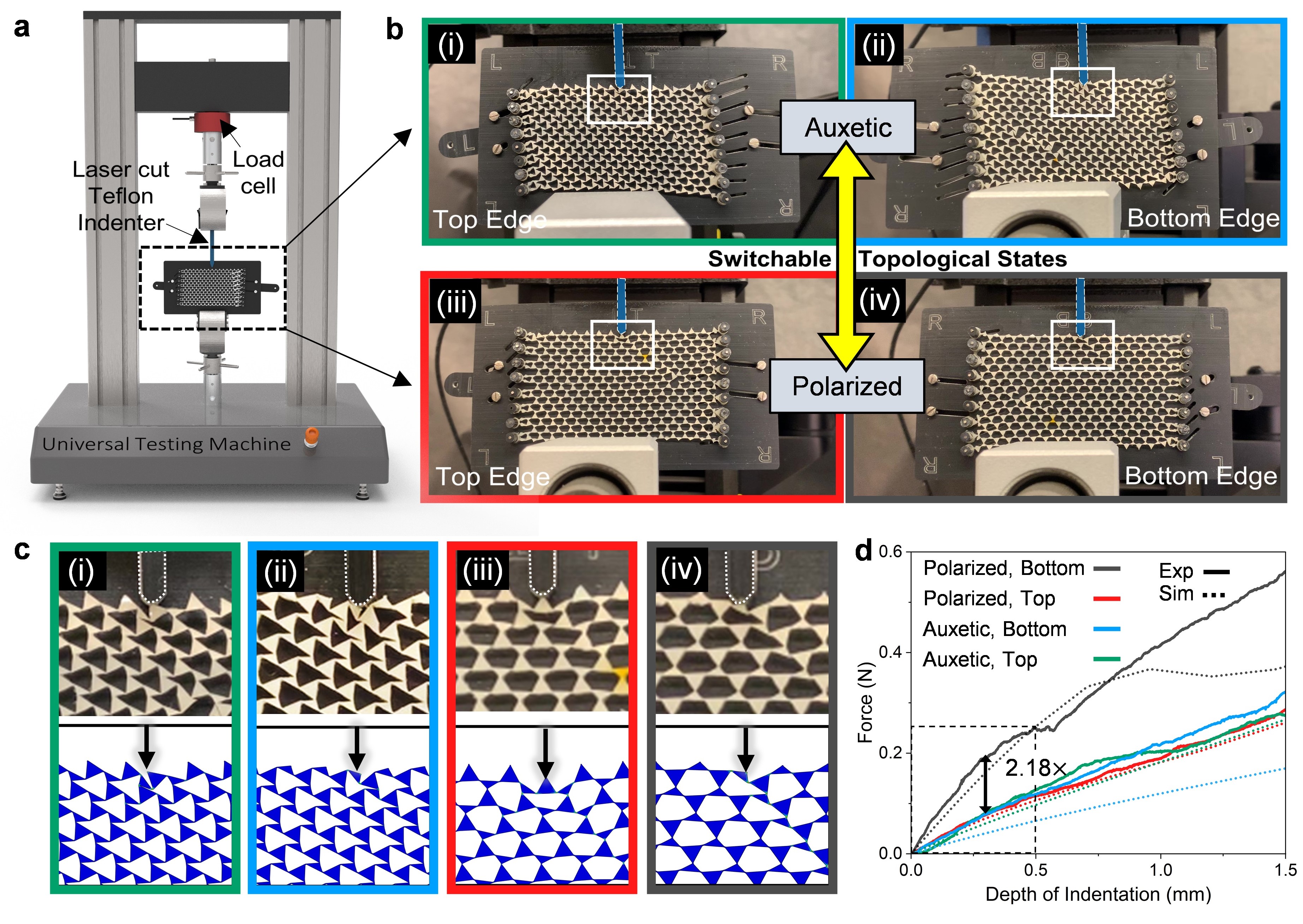}}{}
\makeatother 
\caption{{Experimental characterization of topological edge behavior by indentation. a) Schematic of the experimental setup to probe the mechanical edge response of the TTMM. b) Optical images showing local invaginations of the (i,iii) top and (ii,iv) bottom edges of the lattice in its (i,ii) auxetic and (ii,iv) polarized states, respectively. c) Simulated indentations of auxetic and polarized lattice edges juxtaposed with equivalent experimental results. d) Force-displacement (f-d) curves from edge indentation tests of auxetic and polarized lattices. PCLDA-SMP TTMM samples with 100 $\mu$m hinges are used in these tests.}}
\label{Figure_3}
\end{figure}
\egroup

\subsection{Probing topological edge behavior}
The mechanical response of TTMM lattices subjected to local invagination at the edges is studied both experimentally and via FEM.  The lattice-jig assembly is directly mounted in a universal testing machine outfitted with a sensitive load cell and a custom PTFE indenter (\textbf{Figure~\ref{Figure_3}}a). All lattices are left mounted in their jigs during testing because: (i) The lattices are small, thin and delicate and as such some sort of brace / support is essential; (ii) while the lattice can be left untethered in-plane and sandwiched between two rigid sheets to prevent out-of-plane buckling, we face issues such as high friction and / or loss of line-of-sight of the deformation. Figure~\ref{Figure_3}b and Movie S4, Supporting Information, show snapshots and videos of the indentation of the top (i, iii) and bottom (ii, iv) edges of the same PCLDA-SMP TTMM lattice in both its auxetic (i, ii) and polarized (iii, iv) phases, respectively. Regions of interest where the bulk of the deformation is localized in each test are isolated in Figure~\ref{Figure_3}c and show good agreement with the juxtaposed results from FEM. The slopes of the force-displacement (f-d) curves (i.e., stiffness), measured within the (linear) limit of indentation at various triangular units along the lattice edges, reveal the static or `zero' frequency elastic response of the geometry (see Figure~\ref{Figure_3}d). The experimental data is in good agreement with FEM results quantitatively, validating the elastic-plastic material model in our simulations. Ratios of the slopes of f-d curves (i.e., stiffness ratios, S.R.) obtained by indenting opposite lattice edges quantify a polarized mechanical response (or lack thereof). S.R. has a theoretical minimum value of 1 in the auxetic state but can increase to a maximum value that depends on hinge width when a sample is polarized, which will be discussed later.   

We further investigate the effects of various indenter geometries, edge pinning boundary conditions, lattice sizes and aspect ratios on the probed edge behaviors, so as not to inadvertently introduce any `artificial' S.R. between the opposite edges. For instance, the geometry of the indenter tip can be engineered to suppress stick-slip behavior during indentation but potentially at the cost of suppressing local triangular rotations and measuring an artificially stiffer response. As shown in Figure S7, Supporting Information, three representative indenter tip geometries are considered, including a flat, a small notched and a large, V-shaped grooved tip. As anticipated, the V-shaped tip measures a stiffer response across the board. In experiments, it is observed to disproportionately stiffen the bottom edge of auxetic lattices by simultaneously engaging multiple smaller triangular units and suppressing rotations at the hinges. In contrast, the theoretically-ideal flat tip measures the softest responses and does not suppress local rotations at the site of indentation. However, stick-slip behavior is encountered when the indenter engages the lateral edges of the rotating triangular units. Among the three geometries considered, a tip with a small notch offers a healthy compromise by positively engaging triangular edge units with negligible artificial `polarization' induced in the auxetic lattice (observed as a small difference in the slopes of the top and bottom edges of the auxetic samples). In the polarized configuration, the small notch does not appreciably alter the stiffness of the rigid edge but mildly stiffens the floppy edge as evidenced by the slight reduction in measured S.R.'s. 

The effect of the experimental boundary conditions of (i) leaving the lattice pinned in its kinematic jig during testing as opposed to (ii) leaving it sandwiched between two rigid plates and resting on its opposite edge were simulated. Figure S8, Supporting Information, reveals that at small depths of indentation (up to 0.5 mm), the mechanical response is identical but begins to diverge as indentation depth increases, especially in the polarized configuration: as the region affected by the local deformation at the indentation site propagates and begins to pull against the additional pinning constraints at the L-R edges. Tests on `narrower' lattices, with fewer units along the `\textit{x}' direction and pinned boundaries closer to the site of indentation, further supports this assessment. Therefore, we limit calculations of edge stiffness to be within a 0.5 mm indentation depth. The impact of various lattice sizes and aspect ratios on polar elasticity is also investigated. The f-d results are plotted in Figure S9, Supporting Information, with increasing numbers of lattice units along the x- and y-axes. As observed earlier, narrow lattices with increasingly fewer units along the horizontal (x) direction, are imbued with an `artificial' polar response courtesy the pinning constraints at the L-R edges. In contrast, lattices with increasingly fewer rows of units along the vertical (y) direction show increasingly diminished polarization. A critical threshold is reached in lattices with less than 4 rows of units wherein the S.R. suddenly approaches 1. In this work, lattices are standardized to have 16x8 repeating units - which is sufficiently large to capture the bulk behavior of TTMMs while remaining tractable for fabrication and testing.

In order to demonstrate the repeatability and cyclability of our TTMM, we perform 10 complete kinematic cycles of transforming the lattice between its auxetic and polarized states. In each cycle, the sample's mechanical edge behavior is characterized. As seen from Figure S10, Supporting Information, S.R.'s of the polarized ($\sim$2.1x on average) and auxetic ($\sim$1.15x) samples are consistently recovered after each cycle, confirming that topological polar elasticity can indeed be modulated in a TTMM and is robust against sample fatigue.

\bgroup
\fixFloatSize{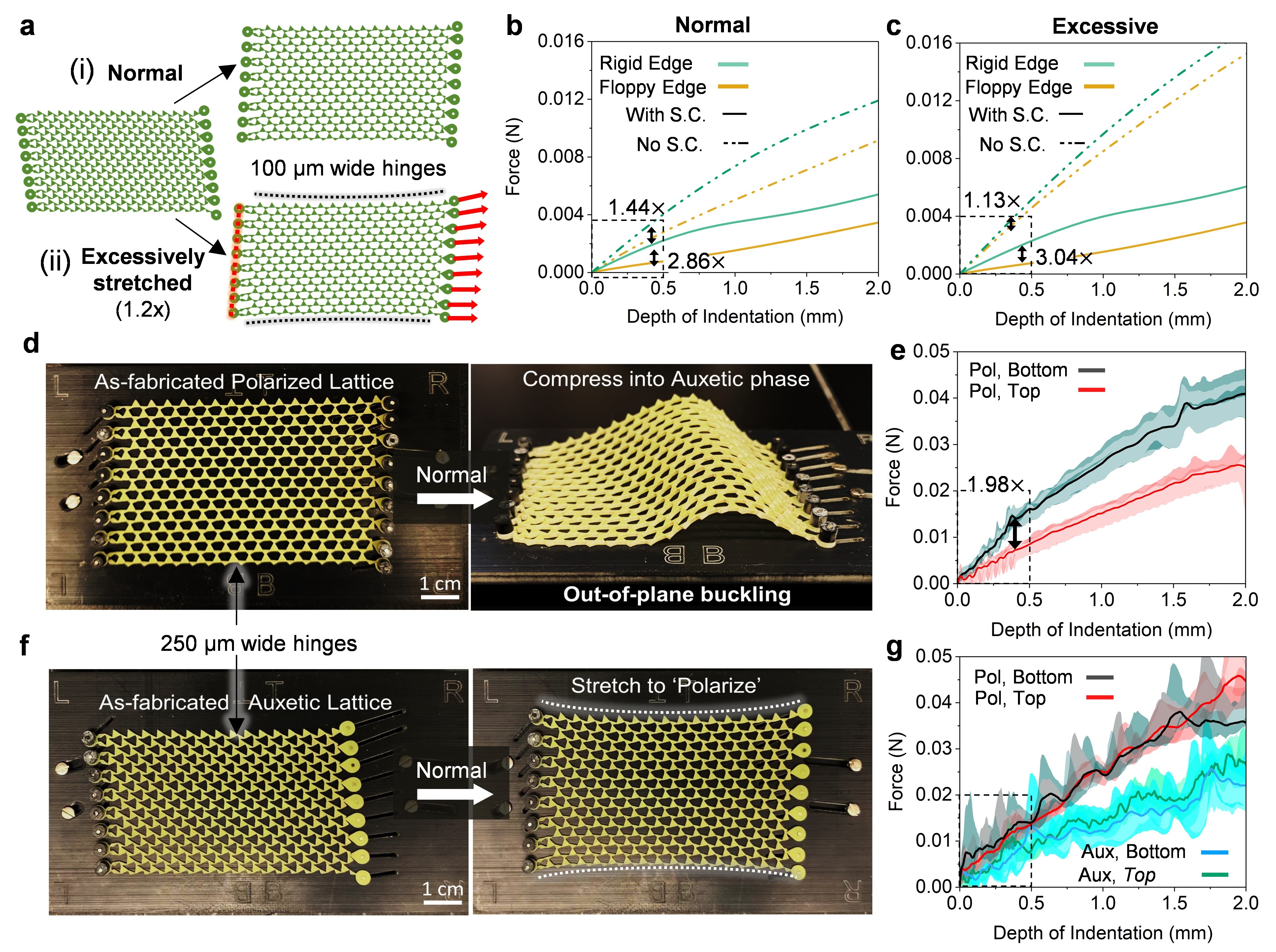}
\begin{figure}[!htbp]
\centering \makeatletter\IfFileExists{Figures/Figure_4.jpg}{\includegraphics{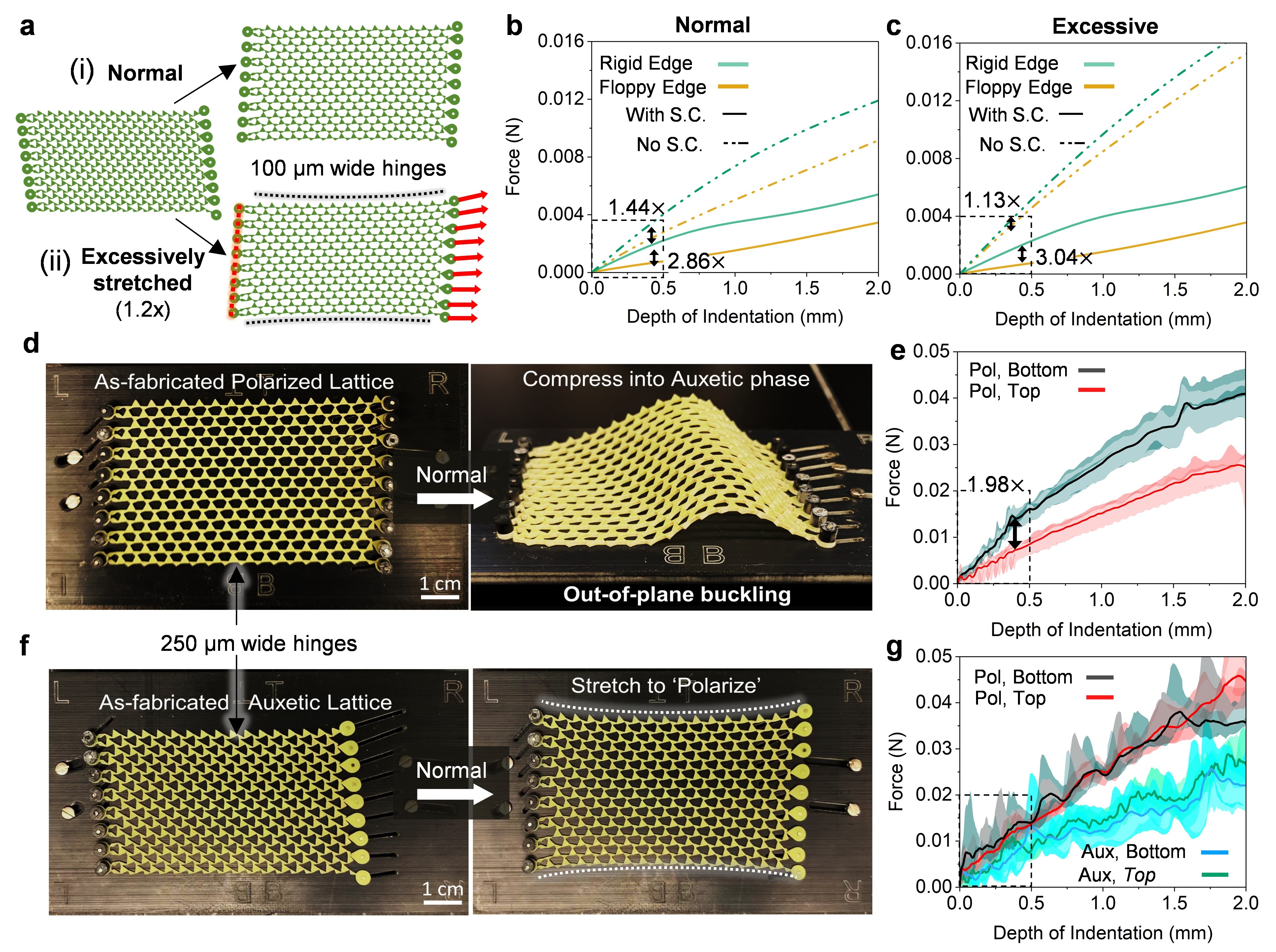}}{}
\makeatother 
\caption{{Stress caching protects topologically polarized mechanical responses from elastic stresses generated during kinematic transformations. a-c) Simulated edge indentation of auxetic lattices with 100 $\mu$m hinges following (i,b) a normal kinematic transformation into the polarized phase, and (ii) an abnormal transformation with 20\% excess stretching of the right edge. Comparing f-d data before and after stress caching (S.C.), reveals a (b) significantly attenuated or (c) destroyed polar edge response in the absence of S.C., that recovers completely post-S.C.. d,f) Reference samples with 250 $\mu$m wide hinges, laser cut in (d) polarized and (f) auxetic configurations, respectively, from a commercial silicone elastomer incapable of S.C.. High in-plane stiffness precluded the transformation of the (d) as-fabricated polarized sample into the auxetic phase. The (f) as-fabricated auxetic lattice could be stretched into the polarized configuration, albeit with noticeable curvature of the top-bottom edges. e,g) Experimental indentation results showing the destroyed topological polarization of the (g) transformed auxetic lattice relative to the (e) as-fabricated polarized sample, due to the large stresses generated, coupled with the lack of S.C..}}
\label{Figure_4}
\end{figure}
\egroup

\subsection{Effects of kinematic stress history on polar elasticity}
Our use of PCLDA-SMP has thus far been primarily motivated by its shape memory effect and stress caching abilities, which stabilize various lattice conformations in the absence of confinement or a continuous mechanical input. However, a closer look at the influence of stored elastic stresses on a topological lattice's behavior reveals that the material's intrinsic ability to modulate polymer chain mobility and cache stresses does more than just stabilize a given lattice conformation: it is also pivotal in preserving its polar elasticity. To support this, we turn to reference samples made from the ED-32 elastomer to study the effects of uncached kinematic stresses. Edge indentations on ED-32 lattices are simulated assuming hyperelastic material properties. As shown in \textbf{Figure~\ref{Figure_4}}a, an auxetic lattice is stretched into its polarized state and its top and bottom edges are probed immediately after. To compare, hypothetical effects of stress caching in these samples are simulated indirectly by indenting the stretched, `deformed' FEM mesh geometry without carrying over any kinematic stress. While the more trivial effect of the stored stress in causing a stiffer response at both edges is immediately obvious from the f-d curves, a more subtle but detrimental effect is observed in the reduction of the S.R. of the stressed polarized lattice, 1.44x compared to 2.86x from the same lattice after erasing its stress history - a nearly 2x increase as a result of stress caching (Figure~\ref{Figure_4}b). This phenomenon is exaggerated when the right edge of the lattice is stretched excessively (20\% more than required for polarization), thereby storing a larger amount of tensile stress in the hinges. Here, the polar response of the stressed lattice is almost entirely annihilated (S.R.$\sim$1.13x) but recovers dramatically by 2.69x (i.e.,S.R.$\sim$3.04x) upon erasing its stress history prior to indentation (Figure~\ref{Figure_4}c). 

These results are validated experimentally in auxetic and polarized lattices laser cut from a 1 mm thick sheet of ED-32 (Figure~\ref{Figure_4}d, f). Due to fabrication and material limitations, these lattices have 250 $\mu$m wide hinges as opposed to the 100 $\mu$m wide hinges in the PCLDA-SMP lattices. A significantly polar response with an S.R.$\sim$1.98x is measured in the as-fabricated polarized ED-32 lattice (Figure~\ref{Figure_4}e). However, courtesy the wider hinges, the lattice is too `stiff' in-plane relative to its out-of-plane compliance, and preferred to buckle instead of transforming into its auxetic phase (Figure~\ref{Figure_4}d). As shown in Figure~\ref{Figure_4}g, the probed mechanical behavior of the as-fabricated auxetic lattice shows no polarization after being stretched into its `polarized' phase. Instead, a uniformly stiffer response of both edges is observed relative to its auxetic behavior. This loss of polarization is attributed to the inability of ED-32 to cache stresses generated in its hinges. A telltale sign of excessive elastic restorative stress in a lattice is the curvature of the top and bottom lattice edges after being stretched into the polarized state (see Figure~\ref{Figure_4}a (ii) and ~\ref{Figure_4}f).

\bgroup
\fixFloatSize{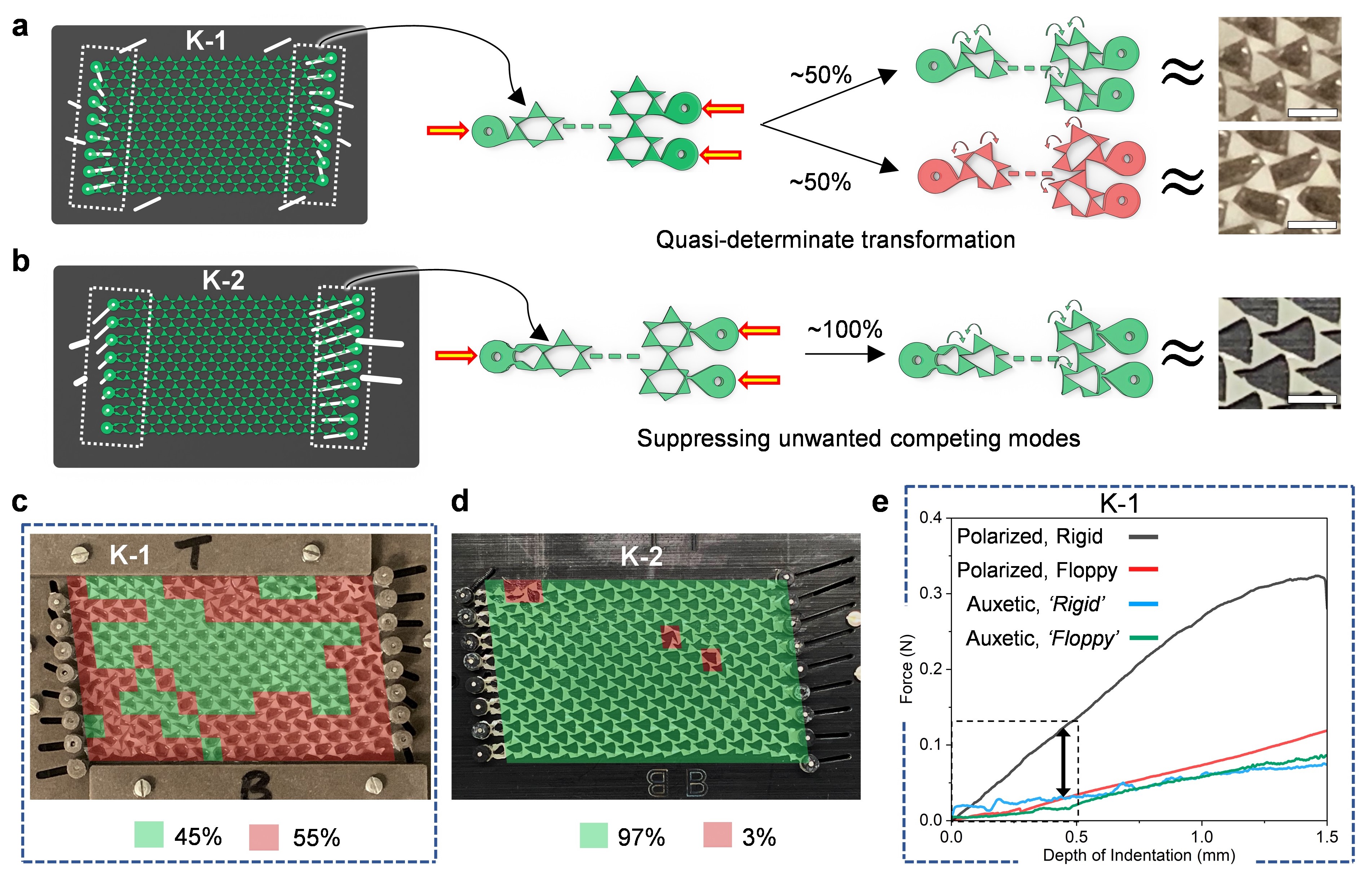}
\begin{figure}[!htbp]
\centering \makeatletter\IfFileExists{Figures/Figure_5.jpg}{\includegraphics{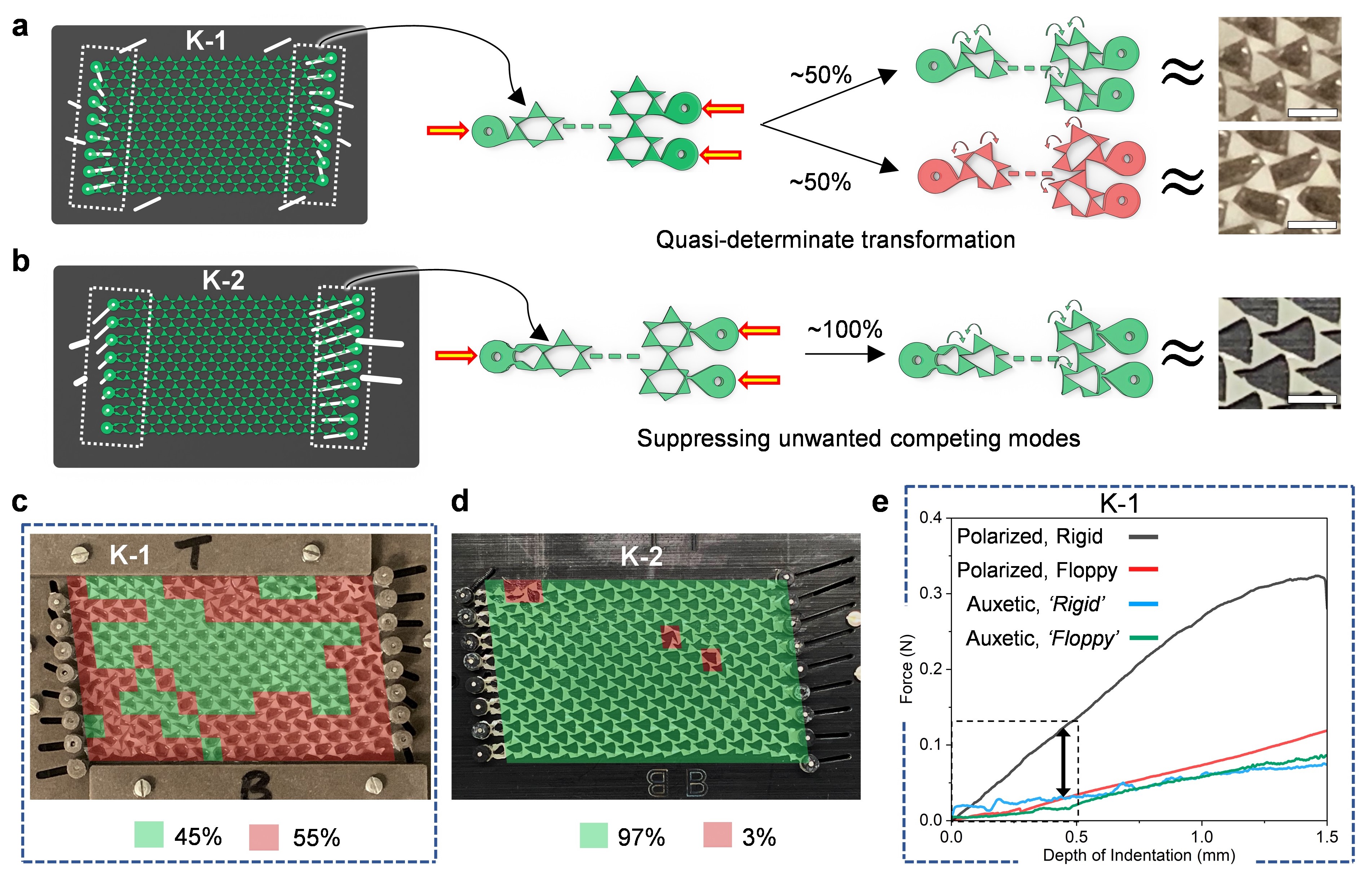}}{}
\makeatother 
\caption{Design of lattice edges to suppress uncertainty in kinematic transformations. a,b) Schematic illustrations of TTMM lattice configurations at topological phase boundaries with edge loop manipulators connected to (a) only regular triangles (K-1) and (b) both triangular sub-units via scissor mechanisms (K-2), respectively. Bulk floppy modes set up by sample-spanning aligned triangle edges (bonds) lead to buckling indeterminacies during compression. a) Exerting kinematic control via half the edge sub-units results in a quasi-determinate transformation into the auxetic state, courtesy closely competing local modes. b)  Undesirable modes can be suppressed by using scissor mechanisms to induce preferential rotations, resulting in a wholly-determinate global transformation. c,d) Optical images of as-fabricated polarized (a) K-1 and (d) K-2 lattices after being compressed into their auxetic state, overlaid with transformation-accuracy ‘heat maps’ that color code local repeating unit conformations as desirable (green) and undesirable (red). e) Experimental edge indentation of K-1 samples reveal that its auxetic behavior is robust against disorder.}
\label{Figure_5}
\end{figure}
\egroup

\subsection{Designing lattice edges to mitigate kinematic indeterminacy}
We note a special case arising at phase boundaries where straight lines of bonds (edges of triangles) align and span the extent of a sample, temporarily transforming surface floppy modes into bulk modes. Consequently, as shown in \textbf{Figure~\ref{Figure_5}}, when a lattice is subjected to compressive vector displacements at its edges, to transform a polarized configuration into the auxetic phase, local buckling indeterminacies impede its compaction into an ordered auxetic lattice. This is encountered experimentally while transforming a polarized TTMM lattice configuration residing exactly on the phase boundary between the first and second polarized phases, with straight lines of horizontally aligned bonds. For a specific lattice design referred to as K-1, that has edge loops connected directly to the corners of the larger of the two triangular sub-units (Figure~\ref{Figure_5}a), a nearly 50\% split between desirable and undesirable local rotations is observed upon its compression i.e., a quasi-determinate transformation (Figure~\ref{Figure_5}c). The design of the K-1 lattice, its kinematic jig and their assembly can be found in Figure S11, Supporting Information. With the addition of right- and left-handed scissor mechanisms at the L-R edges of the optimized lattice design introduced earlier, referred to here as K-2, the unwanted local modes can be suppressed as shown in Figure~\ref{Figure_5}b, leading to a 97\% deterministic transformation (3\% attributed to local broken hinges) into the auxetic phase, where lattice units rotate preferentially due to biased buckling (Figure~\ref{Figure_5}d). It must be noted that despite a high level of disorder in the K-1 lattice's first auxetic phase, its mechanical response is preserved, i.e., the opposite edges are equally soft. The lattice is also able to stretch into a pristine polarized phase every time due to the lack of any competing modes and thus, shows typical, highly polar edge behavior in this state (Figure~\ref{Figure_5}e). Eliminating competing modes can be crucial in preserving topological behaviors during transitions between the first and second polarized state in use cases that explore the entire phase space. Here, we clearly differentiate our proposed method from prior work\unskip~\autocite{Wu2015DirectingUnits}, which uses asymmetric or pre-twisted units to eliminate competition between `collapsing' modes and thereby program maximally-auxetic responses in regular kagome lattices. In contrast, our lattice has to be transformed between both dilation-dominant (auxetic) and shear-dominant (polarized) phases, and exhibits topological behaviors that are strongly coupled to the the twisting angle, $\theta$. Importantly, by modifying the lattice edges with scissor mechanisms, we separate the key geometric parameter, $\theta$, that defines the topological phase space, from the geometric modifications required to suppress buckling indeterminacy at phase boundaries.

\bgroup
\fixFloatSize{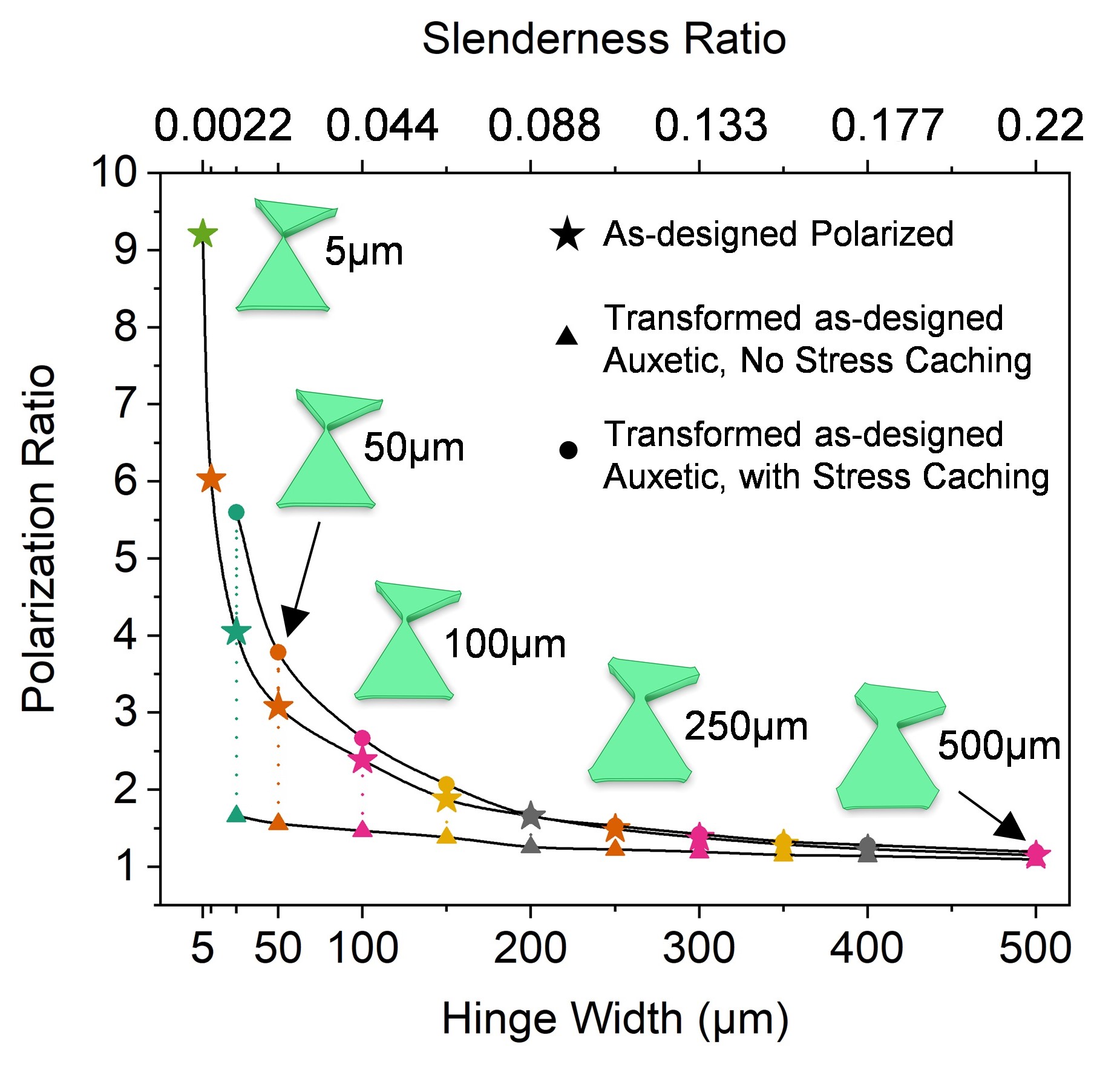}
\begin{figure}[!htbp]
\centering \makeatletter\IfFileExists{Figures/Figure_6.jpg}{\includegraphics{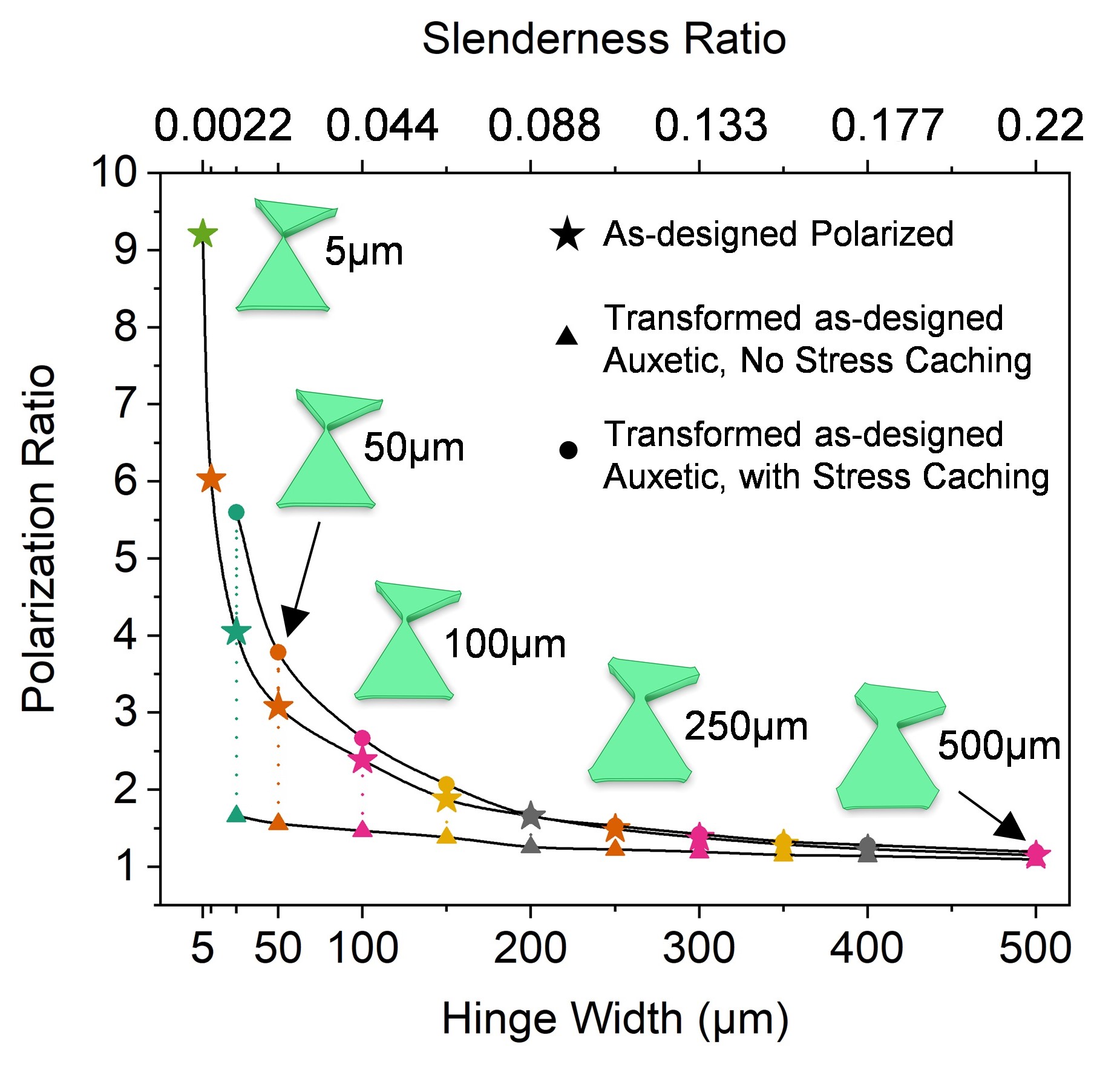}}{}
\makeatother 
\caption{{Effect of hinge slenderness on topological polarization. Simulated edge indentation of lattices with varying hinge widths, reveal that stiffness ratios increases highly non-linearly by nearly an order of magnitude with increasing hinge slenderness, from a saturated minimum value $\sim$1 for the widest hinges considered.}}
\label{Figure_6}
\end{figure}
\egroup

\subsection{Evidence for the topological origins of polarization}
It is evident from simulations that the majority of kinematic stress generated in a lattice is localized in its hinges. The effect of hinge width on the topological polarization exhibited by `as-designed' polarized and kinematically transformed `as-designed' auxetic TTMMs is therefore studied systematically as shown in \textbf{Figure~\ref{Figure_6}} and Figure S12, Supporting Information. Simulated indentations are performed on `as-designed' auxetic lattices that have been stretched into their polarized states and on `as-designed' polarized lattices that serve as reference samples. An intuitive observation is that lattice edges get stiffer with increasing hinge width. However, a more profound realization is that polarization and S.R. increase highly non-linearly with decreasing hinge width or increasing hinge slenderness, as we approach the theoretically-ideal free hinge. This is indirect yet compelling evidence of the topological origins of the polar elasticity observed in this work, as opposed to being a trivial consequence of geometric asymmetries at the lattice edges, and complements prior research on the effects of adding next nearest neighbor bonds on topological modes\unskip~\autocite{Stenull2019SignaturesLattices} and parametric investigations into the effects of hinge width on wave transport at finite frequencies\unskip~\autocite{Ma2019InfluenceStudy}. Our results appear to be material independent as nearly identical results are obtained from simulations with hyperelastic (ED-32) and elastic-plastic (PCLDA-SMP) material properties (Figures S13 and S14, Supporting Information).

\section{Conclusion}
In summary, we demonstrate for the first time, experimental modulation of topological polar elasticity in a monolithic metamaterial made from a shape memory polymer. We establish a kinematic strategy to program reversible global transformations between two topologically distinct phases in a high degree of freedom lattice with a Guest-Hutchinson mode, only via sample edges. We achieve this with only two simple uniaxial mechanical inputs, through a confluence of vector edge mapping, modified lattice edge units and a novel jig design. The material's intrinsic shape memory properties play a key role in stabilizing each lattice conformation without the need for a continuous input. We corroborate the robustness of topologically protected polar edge behavior against defects and sample fatigue by reversibly switching the metamaterial over 10 cycles. Our work reveals that topological polar elasticity is highly sensitive to the kinematic stress history of a transformable metamaterial and demonstrates how polymer chain mobility can be modulated to lock away generated stresses and protect this behavior. Finally, we provide indirect evidence of the topological nature of polar elasticity in lattices with `real' hinges by systematically studying the effects of varying hinge width. This work lays the groundwork for future realizations of reconfigurable devices capable of switching non-trivial and defect tolerant topological behavior for applications such as switchable acoustic diodes\unskip~\autocite{Zhou2019}, tunable vibration dampers or isolators\unskip~\autocite{Kadic2013MetamaterialsElectromagnetism} and tires capable of adapting to diverse terrains\unskip~\autocite{Zunker2021SoftWheels}.

\section{Experimental Section}

\subsection{Synthesis of Polycaprolactone diacrylate (PCLDA)}
PCLDA was synthesized according to the literature\unskip~\autocite{Zhao2016}. Specifically, 50 g of polycaprolactone flakes (PCL, M\ensuremath{_{n}} = 10,000, Sigma Aldrich) was added to 250 g of toluene (Thermo Scientific) in a round bottomed reaction flask that was then heated to 55 $^\circ$C in a silicone oil bath and stirred using a magnetic stir bar for 15 minutes until the PCL was completely dissolved, following which the reaction flask was then lifted out of the oil bath but kept stirring. 3.233 g of triethylamine (TCI America) and 2.716 g of acryloyl chloride (Sigma Aldrich) were measured out in separate vials and added to the flask in order, very slowly and in a drop-wise manner. The reaction mixture turned cloudy and was left stirring at 60 $^\circ$C for 24 h. The flask was then removed from the oil bath and the reaction products were filtered through a 0.22 $\mu$m polyvinylidene fluoride membrane filter (PVDF, 47 mm Durapore, Sigma Aldrich) under vacuum, using a ceramic funnel, rubber adaptor and vacuum Erlenmeyer flask setup, to separate the unwanted salt byproduct. The filtrate was then slowly poured into a beaker containing 800 mL of cold methanol (Fisher Chemical), while stirring constantly using a glass rod. The precipitated solid was filtered out and rinsed a few times with fresh methanol to remove any residual toluene, followed by vacuum filtration through a cellulose filter paper (Whatman Grade 1) several times. The obtained off-white powder was collected and vacuum dried at room temperature overnight. This dried powder was used as-is. 

\subsection{Characterization of intrinsic properties of the PCLDA-based shape memory polymer, PCLDA-SMP}
The melting temperature (T\ensuremath{_{m}}) was characterized via differential scanning calorimetry (TA Instruments DSC Q2000). Sample temperature was scanned from -10 $^\circ$C to 100 $^\circ$C with a ramping rate of 10 $^\circ$C/min  under N\ensuremath{_{2}}. The elastic modulus of the bulk polymer was obtained from tensile tests performed according to the ASTM D638 standard, using dog bone specimens. The polymer's shape memory effect was characterized by a dynamic mechanical analyzer (TA, Q800) in the controlled force mode. A PCLDA-SMP sample (3 cm x 3 cm x 0.3 mm) heated above its T\ensuremath{_{m}} was subjected to an applied stress of 0.3 MPa and then cooled below its crystallization temperature. At this point the applied stress was removed and the percentage of strain retention was observed. Upon reheating above T\ensuremath{_{m\ }}, the sample returned to its original zero strain state, marking the end of one complete cycle. This process was repeated for at least three cycles with {\texttildeapprox}99.8\% shape fixity, indicating a robust shape memory performance.  To characterize the material's stress caching abilities, a similar sample was tested under the instrument's strain rate mode, with an applied strain of 105\%. The temperature ramping rate was fixed at 10 $^\circ$C/min.

\subsection{Fabrication of the metamaterial lattices}
The PCLDA-SMP lattices were fabricated via a multi-step process (see Figure S5, Supporting Information) to reliably manufacture samples with up to 128 mm-scale repeating units connected by microscale hinge ligaments. Specifically, 15g of PCLDA powder, 0.386 g of pentaerythritol tetrakis(3-mercaptopropionate) crosslinker (PETMP, Sigma Aldrich) and 3.86 g of 1 wt\% solution of 2,2-dimethoxy-2-phenylacetophenone photoinitiator (DMPA, Sigma Aldrich) in toluene, were added to a glass vial (Figure S5a, Supporting Information). The vial was capped and the mixture was heated to 75 $^\circ$C (above  T\ensuremath{_{m}}) in an oven to melt the PCLDA powder. The mixture was stirred vigorously to homogenize the components followed by degassing in a vacuum oven for 10 min. A multilayer substrate was used in the fabrication process as shown in Figure S5b, Supporting Information, comprising (from top to bottom) an unpolished 6" silicon wafer as a micro-rough contacting surface, a PTFE sheet that allowed easy removal / peeling off of any crosslinked overflow and a glass sheet as a generic support layer / backbone. The thickness of the cast film was controlled at 1 mm using a high-temperature silicone rubber spacer (Mcmaster Carr) attached to the unpolished side of a Si substrate using double-sided adhesive tape (Scotch Permanent Double Sided Tape, 3M). The degassed molten viscous PCLDA-SMP precursor was cast onto the Si wafer and degassed at 75 $^\circ$C in a vacuum oven for 1 h. A chrome photomask was brought into soft contact with the silicone spacer and the molten precursor without trapping any air bubbles. The cast film was then exposed to 160 mJ/cm\ensuremath{^{2}} of 365nm UV light for 8s (Newport model 97436-1000-1, Hg source) to pattern the lattice. The exposed sample was allowed to cool to room temperature and crystallize before being submerged in a bath of cold isopropanol (IPA, Fisher Chemical) for 12h. At that point, the photomask self-delaminated from the underlying PCLDA-SMP film, and silicone spacer was subsequently removed. The sample was left to soak in IPA for another 24h - 48h to facilitate the complete separation of the lattice from the underlying Si substrate. The patterned sheet of PCLDA-SMP was then developed in a bath of hot toluene (60 $^\circ$C) for 1 h to completely dissolve all the unexposed / uncrosslinked PCLDA. The crosslinked lattice would swell significantly, become transparent and curl up into a cylinder. It was then removed from the hot toluene bath and rinsed in a room temperature toluene bath. The lattice was then flattened and sandwiched between two PTFE mesh sheets (Stretchable high-temperature PTFE plastic mesh 0.045" x 0.025" openings, McMaster Carr). The residual toluene in the developed lattice sample was then evaporated overnight in an organic solvent vacuum oven at 75 $^\circ$C. The sample was cooled to room temperature and the lattice with its edge loops was liberated from its `fabrication frame' by cutting along predesigned cutting lines using a sharp X-acto knife. Care was taken to not apply excessive tensile forces on the lattice hinges during this process. The liberated lattice was then mounted in a kinematic jig via its edge loops for transformation and testing.

\subsection{Kinematic Transformation}
The PTFE jigs (as shown in Figures S7 and S8, Supporting Information) used to prescribe the kinematic transformation of the TTMM lattices were laser cut from 1/8" thick chemical-resistant slippery Teflon{\textsuperscript{\textregistered}} PTFE sheets (McMaster Carr) to minimize friction between the lattice and the surface of the jig and assembled using 2-56 screws and knurled thumb nuts. By compensating for laser kerf, the vector slots were laser cut in the PTFE jig with sufficiently tight tolerances to ensure a sliding clearance fit. The lattices were mounted in their jigs using 1 mm diameter stainless steel linear motion shafts (McMaster Carr) that were first cut down to size using a handheld Dremel rotary tool equipped with a 420 cut-off wheel. Both ends of each linear rod were capped off by laser cut, friction fit `end caps'. L-R slotted guides convey solitary applied mechanical inputs at each edge to the linear rods while accommodating their relative sliding as the lattice dilates in the auxetic phase and then shears in the polarized phase. The guides are outfitted with locking screws to temporarily fix a given lattice configuration, especially during the heating and cool down stages of the transformation process. Most lattice samples were fabricated in their auxetic configurations and as-such, the first kinematic transformation involved heating the samples above their T\ensuremath{_{m}} and `stretching' them into their topologically polarized configuration. This stretching step was broken down into two distinct displacements ($u_x^{L}$ and $u_x^{R}$) applied to the left and right guides, respectively. While these displacements could be applied simultaneously, they were instead usually prescribed sequentially for ease of manual operation, with care being taken to displace the left edge first followed by the right edge during the auxetic-to-polarized transformation. Note that this sequence is reversed during the polarized-to-auxetic `compression' transformation. This was to ensure the lattices were not stretched excessively which would have caused their slender 100 $\mu$m hinge ligaments to either fail plastically or catastrophically. At times, once the target-phase-specific edge unit positions had been prescribed using the jig, local conformational defects would arise, due to local stiction between the intrinsically soft (above its T\ensuremath{_{m}}) and extrinsically compliant PCLDA-SMP lattice and the smooth jig surface, despite the use of a low surface energy material such as PTFE and even lubricating silicone oil (Polydimethylsiloxane, Thermo Scientific\textsuperscript{TM} Catalogue No. AC163850025, viscosity = 500 mPa.s at 25 $^\circ$C). Fortunately such defects could be corrected quite easily with a gentle nudge, allowing the stuck lattice unit to `relax' to its equilibrium position. The transformed lattices were allowed to cool to room temperature prior to mechanical testing.

\subsection{Static Mechanical Testing}
Edge stiffness of the PCLDA-SMP TTMM lattices was measured from quasi-static indentation using an Instron{\textsuperscript{\textregistered}} Model 68SC-2 universal testing machine equipped with a Model 2350-50N load cell. A custom laser-cut PTFE indenter was held in a Model 2716-016 manual wedge action grip connected to the load cell. The lattice + kinematic jig was held in a Model 2710-113 screw side action grip. Indentation tests on laser cut Elite Double 32 lattices were performed using an Instron{\textsuperscript{\textregistered}} Model 5564 equipped with a 2.5N load cell (Model 2525-815). The PTFE indenter was held in Model 2712-101 micro-pneumatic side action grips connected to the load cell and the lattice-kinematic jig assembly was held in Model 2712-020 pneumatic side action grips. All tests were performed under displacement control up to a maximum indentation depth of 2 mm at a rate of 2 mm/min. The force-displacement (f-d) data was appropriately truncated and re-normalized to compensate for baseline loads associated with friction and initial self-truing of minor misalignments between the indenter and the jig. All values of edge stiffness and S.R.s were measured from F-d data corresponding to indentation depths less than 0.5 mm.

\subsection{Finite Element Modeling} 
FEM simulations were performed in ABAQUS 6.24/CAE 2020 (Dassault Syst{\`e}mes\ensuremath{^{{\textsuperscript{\textregistered}}}} Simulia Corp.) using ABAQUS/Standard. The properties of the PCLDA-SMP polymer were described by an elastic-plastic material model with Young's modulus, E = 132 MPa and a Poisson's ratio, $\nu $= 0.48. Post-yield stress-strain data from tensile tests was included in the material model. We model the mechanical response of the ED-32 elastomer using an incompressible Gent model\unskip~\autocite{Gent1996ARubber} with a strain energy density function \textit{W} given by,
\begin{eqnarray*}W=-\frac{\mu J_{lim}}2\ln\left(1-\frac{I_1-3}{J_{lim}}\right) \end{eqnarray*}
where $\mu $ represents the small strain shear modulus, $J_{lim} $ is a material parameter related to the limiting value of stretch and $I_1=tr\;\left(\mathbf F^\mathbf T\mathbf F\right) $, where $\mathbf{F}$ is the deformation gradient. An ABAQUS user subroutine (UHYPER) described in the literature\unskip~\autocite{Jin2020Kirigami-InspiredShapes} was then used to define the hyperelastic behavior of ED-32 in our simulations. All simulation material parameters were informed by tensile test data (Figure S2, Supporting Information). All lattice variants were discretized with CPS4 4-node bilinear plane stress quadrilateral elements and indented under displacement control. `Hard' surface-to-surface contact with a nominal friction coefficient value of 0.3 and separation after contact being allowed, precluded any local penetration / overlap of mesh elements. As far as possible, lattice geometries were seeded with a sufficiently fine local mesh so as to have at least three mesh elements span the width of their finest features i.e., the 5$\mu$m - 500$\mu$m wide hinges.

\section*{Acknowledgments}
The work is supported by the National Science Foundation (NSF) Emerging Frontiers in Research and Innovation (EFRI) NewLAW grant, \#1741618. The authors also acknowledge the use of SEM resources supported by the NSF/Materials Research Science and Engineering Center (MRSEC) at the University of Pennsylvania, \#DMR-1720530. K.S. and X.M. also acknowledge the support of the Office of Naval Research, MURI grant \#N00014-20-1-2479. We thank Tom Lubensky for his invaluable insight and feedback. We thank Mohammad Charara for several helpful discussions. We thank Steven Szewczyk for his valuable assistance with various experiments. We also thank Jiaqi Liu and Mingzhu Liu for their assistance with SEM. 

\section*{Author Contributions}
J.C.J., X.M., B.J. and S.Y. conceived the idea. J.C.J. and B.J. contributed equally to this work. J.C.J. and B.J. synthesized PCLDA and PCLDA-SMP and characterized their mechanical behavior. J.C.J. fabricated samples and collected data. J.C.J. performed the FEM simulations with assistance from L.J and Y.L. J.C.J. analyzed the results. S.Y. and T.X. supervised the research. J.C.J. and S.Y. wrote the manuscript. All authors discussed the results. 

\section*{Conflict of Interest}
The authors declare no conflict of interest.

\printbibliography

@article{Bilal2017a,
    title = {{Bistable metamaterial for switching and cascading elastic vibrations}},
    year = {2017},
    journal = {Proceedings of the National Academy of Sciences of the United States of America},
    author = {Bilal, Osama R. and Foehr, André and Daraio, Chiara},
    number = {18},
    month = {5},
    pages = {4603},
    volume = {114},
    publisher = {National Academy of Sciences},
    doi = {10.1073/pnas.1618314114},
    issn = {10916490},
    pmid = {28416663}
}

@article{Ma2018,
    title = {{Edge Modes and Asymmetric Wave Transport in Topological Lattices: Experimental Characterization at Finite Frequencies}},
    year = {2018},
    journal = {Physical Review Letters},
    author = {Ma, Jihong and Zhou, Di and Sun, Kai and Mao, Xiaoming and Gonella, Stefano},
    number = {9},
    month = {5},
    volume = {121},
    url = {http://arxiv.org/abs/1805.02759 http://dx.doi.org/10.1103/PhysRevLett.121.094301},
    issn = {10797114},
    pmid = {30230879},
    arxivId = {1805.02759}
}

@article{Bilal2017,
    title = {{Intrinsically Polar Elastic Metamaterials}},
    year = {2017},
    journal = {Advanced Materials},
    author = {Bilal, Osama R. and S{\"{u}}sstrunk, Roman and Daraio, Chiara and Huber, Sebastian D.},
    number = {26},
    month = {7},
    pages = {1700540},
    volume = {29},
    publisher = {John Wiley {\&} Sons, Ltd},
    url = {http://doi.wiley.com/10.1002/adma.201700540},
    doi = {10.1002/adma.201700540},
    issn = {15214095}
}

@article{Cho2019,
    title = {{Intrinsically reversible superglues via shape adaptation inspired by snail epiphragm}},
    year = {2019},
    journal = {Proceedings of the National Academy of Sciences of the United States of America},
    author = {Cho, Hyesung and Wu, Gaoxiang and Jolly, Jason Christopher and Fortoul, Nicole and He, Zhenping and Gao, Yuchong and Jagota, Anand and Yang, Shu},
    number = {28},
    pages = {13774},
    volume = {116},
    publisher = {National Academy of Sciences},
    doi = {10.1073/pnas.1818534116},
    issn = {10916490}
}

@article{Baardink2017,
    title = {{Localizing softness and stress along loops in three-dimensional topological metamaterials}},
    year = {2017},
    journal = {Proceedings of the National Academy of Sciences of the United States of America},
    author = {Baardink, Guido and Souslov, Anton and Paulose, Jayson and Vitelli, Vincenzo},
    number = {3},
    month = {1},
    pages = {489},
    volume = {115},
    publisher = {National Academy of Sciences},
    url = {www.pnas.org/cgi/doi/10.1073/pnas.1713826115 http://www.ncbi.nlm.nih.gov/pubmed/29284745 http://www.pubmedcentral.nih.gov/articlerender.fcgi?artid=PMC5776976 http://arxiv.org/abs/1707.08928%0Ahttp://dx.doi.org/10.1073/pnas.1713826115},
    isbn = {0780358902},
    doi = {10.1073/pnas.1713826115},
    issn = {0027-8424},
    pmid = {29284745},
    arxivId = {1707.08928}
}

@article{Pellegrino1986,
    title = {{Matrix analysis of statically and kinematically indeterminate frameworks}},
    year = {1986},
    journal = {International Journal of Solids and Structures},
    author = {Pellegrino, S. and Calladine, C. R.},
    number = {4},
    month = {1},
    pages = {409},
    volume = {22},
    publisher = {Pergamon},
    doi = {10.1016/0020-7683(86)90014-4},
    issn = {00207683}
}

@article{Mao2018,
    title = {{Maxwell Lattices and Topological Mechanics}},
    year = {2018},
    journal = {Annual Review of Condensed Matter Physics},
    author = {Mao, Xiaoming and Lubensky, Tom C.},
    number = {1},
    month = {3},
    pages = {413},
    volume = {9},
    url = {http://www.annualreviews.org/doi/10.1146/annurev-conmatphys-033117-054235},
    doi = {10.1146/annurev-conmatphys-033117-054235},
    issn = {1947-5454}
}

@article{Rocklin2016,
    title = {{Mechanical Weyl Modes in Topological Maxwell Lattices}},
    year = {2016},
    journal = {Physical Review Letters},
    author = {Rocklin, D. Zeb and Chen, Bryan Gin Ge and Falk, Martin and Vitelli, Vincenzo and Lubensky, T. C.},
    number = {13},
    month = {4},
    pages = {135503},
    volume = {116},
    publisher = {American Physical Society},
    url = {https://journals-aps-org.proxy.library.upenn.edu/prl/abstract/10.1103/PhysRevLett.116.135503},
    doi = {10.1103/PhysRevLett.116.135503},
    issn = {10797114},
    arxivId = {1510.04970}
}

@article{Lubensky2015,
    title = {{Phonons and elasticity in critically coordinated lattices}},
    year = {2015},
    journal = {Reports Progress in Physics},
    author = {Lubensky, T. C. and Kane, C. L. and Mao, Xiaoming and Souslov, A. and Sun, Kai},
    number = {7},
    month = {7},
    volume = {78},
    publisher = {Institute of Physics Publishing},
    arxivId = {1503.01324}
}

@article{Zhao2016,
    title = {{Shape memory polymer network with thermally distinct elasticity and plasticity}},
    year = {2016},
    journal = {Science Advances},
    author = {Zhao, Qian and Zou, Weike and Luo, Yingwu and Xie, Tao},
    number = {1},
    month = {1},
    volume = {2},
    publisher = {American Association for the Advancement of Science},
    issn = {23752548}
}

@article{Pishvar2020,
    title = {{Soft Topological Metamaterials with Pronounced Polar Elasticity in Mechanical and Dynamic Behaviors}},
    year = {2020},
    journal = {Physical Review Applied},
    author = {Pishvar, Maya and Harne, Ryan L.},
    number = {4},
    month = {10},
    pages = {044034},
    volume = {14},
    publisher = {American Physical Society},
    url = {https://journals.aps.org/prapplied/abstract/10.1103/PhysRevApplied.14.044034},
    doi = {10.1103/PhysRevApplied.14.044034},
    issn = {23317019}
}

@article{Sun2012,
    title = {{Surface phonons, elastic response, and conformal invariance in twisted kagome lattices}},
    year = {2012},
    journal = {Proceedings of the National Academy of Sciences of the United States of America},
    author = {Sun, Kai and Souslov, Anton and Mao, Xiaoming and Lubensky, T. C.},
    number = {31},
    month = {7},
    pages = {12369},
    volume = {109},
    publisher = {PNAS},
    doi = {10.1073/pnas.1119941109},
    issn = {00278424},
    pmid = {22733727}
}

@article{Zhou2019,
    title = {{Switchable phonon diodes using nonlinear topological Maxwell lattices}},
    year = {2020},
    journal = {Physical Review B},
    author = {Zhou, Di and Ma, Jihong and Sun, Kai and Gonella, Stefano and Mao, Xiaoming},
    number = {10},
    month = {8},
    volume = {101},
    url = {http://arxiv.org/abs/1908.05716},
    issn = {24699969},
    arxivId = {1908.05716}
}

@article{Kane2013,
    title = {{Topological boundary modes in isostatic lattices}},
    year = {2013},
    journal = {Nature Physics},
    author = {Kane, C. L. and Lubensky, T. C.},
    number = {1},
    month = {1},
    pages = {39},
    volume = {10},
    publisher = {Nature Publishing Group},
    url = {http://www.nature.com/articles/nphys2835 www.nature.com/naturephysics},
    doi = {10.1038/nphys2835},
    issn = {17452481},
    pmid = {20524887},
    arxivId = {1308.0554}
}

@article{Rocklin2017,
    title = {{Transformable topological mechanical metamaterials}},
    year = {2017},
    journal = {Nature Communications},
    author = {Rocklin, D. Zeb and Zhou, Shangnan and Sun, Kai and Mao, Xiaoming},
    month = {1},
    volume = {8},
    publisher = {Nature Publishing Group},
    issn = {20411723},
    arxivId = {1510.06389}
}

@article{Gent1996ARubber,
    title = {{A new constitutive relation for rubber}},
    year = {1996},
    journal = {Rubber Chemistry and Technology},
    author = {Gent, A. N.},
    number = {1},
    month = {3},
    pages = {59},
    volume = {69},
    publisher = {Allen Press},
    url = {https://meridian.allenpress.com/rct/article/69/1/59/92301/A-New-Constitutive-Relation-for-Rubber},
    doi = {10.5254/1.3538357},
    issn = {00359475}
}

@article{Susstrunk2016ClassificationMetamaterials,
    title = {{Classification of topological phonons in linear mechanical metamaterials}},
    year = {2016},
    journal = {Proceedings of the National Academy of Sciences of the United States of America},
    author = {S{\"{u}}sstrunk, Roman and Huber, Sebastian D.},
    number = {33},
    month = {8},
    pages = {E4767},
    volume = {113},
    publisher = {National Academy of Sciences},
    doi = {10.1073/pnas.1605462113},
    issn = {10916490},
    arxivId = {1604.01033}
}

@article{Wu2015DirectingUnits,
    title = {{Directing the deformation paths of soft metamaterials with prescribed asymmetric units}},
    year = {2015},
    journal = {Advanced Materials},
    author = {Wu, Gaoxiang and Cho, Yigil and Choi, In Suk and Ge, Dengteng and Li, Ju and Han, Heung Nam and Lubensky, Tom and Yang, Shu},
    number = {17},
    month = {5},
    pages = {2747},
    volume = {27},
    publisher = {John Wiley {\&} Sons, Ltd},
    url = {https://onlinelibrary-wiley-com.proxy.library.upenn.edu/doi/full/10.1002/adma.201500716 https://onlinelibrary-wiley-com.proxy.library.upenn.edu/doi/abs/10.1002/adma.201500716 https://onlinelibrary-wiley-com.proxy.library.upenn.edu/doi/10.1002/adma.2015007},
    doi = {10.1002/adma.201500716},
    issn = {15214095},
    pmid = {25808041}
}

@article{Meeussen2016GearedStability,
    title = {{Geared topological metamaterials with tunable mechanical stability}},
    year = {2016},
    journal = {Physical Review X},
    author = {Meeussen, Anne S. and Paulose, Jayson and Vitelli, Vincenzo},
    number = {4},
    month = {11},
    pages = {041029},
    volume = {6},
    publisher = {American Physical Society},
    url = {https://journals.aps.org/prx/abstract/10.1103/PhysRevX.6.041029},
    doi = {10.1103/PhysRevX.6.041029},
    issn = {21603308},
    arxivId = {1602.08769}
}

@inproceedings{Ma2019InfluenceStudy,
    title = {{Influence of hinge stiffness on the asymmetric wave transport in topological lattices: a parametric study}},
    year = {2019},
    booktitle = {Health Monitoring of Structural and Biological Systems XIII. Vol. 10972.},
    author = {Ma, Jihong and Zhou, Di and Sun, Kai and Mao, Xiaoming and Gonella, Stefano},
    number = {1},
    month = {4},
    pages = {40},
    volume = {10972},
    publisher = {SPIE},
    isbn = {9781510625990},
    doi = {10.1117/12.2514193},
    issn = {1996756X}
}

@article{Jin2020Kirigami-InspiredShapes,
    title = {{Kirigami-Inspired Inflatables with Programmable Shapes}},
    year = {2020},
    journal = {Advanced Materials},
    author = {Jin, Lishuai and Forte, Antonio Elia and Deng, Bolei and Rafsanjani, Ahmad and Bertoldi, Katia},
    number = {33},
    month = {8},
    pages = {2001863},
    volume = {32},
    publisher = {John Wiley {\&} Sons, Ltd},
    url = {https://onlinelibrary.wiley.com/doi/full/10.1002/adma.202001863 https://onlinelibrary.wiley.com/doi/abs/10.1002/adma.202001863 https://onlinelibrary.wiley.com/doi/10.1002/adma.202001863},
    doi = {10.1002/adma.202001863},
    issn = {15214095},
    pmid = {32627259},
    arxivId = {2007.07312}
}

@article{Maxwell1864L.Frames,
    title = {{L. On the calculation of the equilibrium and stiffness of frames}},
    year = {1864},
    journal = {The London, Edinburgh, and Dublin Philosophical Magazine and Journal of Science},
    author = {Maxwell, J. Clerk},
    number = {182},
    pages = {294},
    volume = {27},
    publisher = {Taylor {\&} Francis},
    url = {https://doi.org/10.1080/14786446408643668},
    doi = {10.1080/14786446408643668},
    issn = {1941-5982}
}

@article{Li2021Liquid-inducedMicrostructures,
    title = {{Liquid-induced topological transformations of cellular microstructures}},
    year = {2021},
    journal = {Nature},
    author = {Li, Shucong and Deng, Bolei and Grinthal, Alison and Schneider-Yamamura, Alyssha and Kang, Jinliang and Martens, Reese S. and Zhang, Cathy T. and Li, Jian and Yu, Siqin and Bertoldi, Katia and Aizenberg, Joanna},
    number = {7854},
    month = {4},
    pages = {386},
    volume = {592},
    publisher = {Nature Publishing Group},
    url = {https://www.nature.com/articles/s41586-021-03404-7},
    doi = {10.1038/s41586-021-03404-7},
    issn = {14764687},
    pmid = {33854248}
}

@article{Montgomery2021Magneto-MechanicalBandgaps,
    title = {{Magneto-Mechanical Metamaterials with Widely Tunable Mechanical Properties and Acoustic Bandgaps}},
    year = {2021},
    journal = {Advanced Functional Materials},
    author = {Montgomery, S. Macrae and Wu, Shuai and Kuang, Xiao and Armstrong, Connor D. and Zemelka, Cole and Ze, Qiji and Zhang, Rundong and Zhao, Ruike and Qi, H. Jerry},
    number = {3},
    month = {1},
    pages = {2005319},
    volume = {31},
    publisher = {John Wiley {\&} Sons, Ltd},
    url = {https://onlinelibrary.wiley.com/doi/full/10.1002/adfm.202005319 https://onlinelibrary.wiley.com/doi/abs/10.1002/adfm.202005319 https://onlinelibrary.wiley.com/doi/10.1002/adfm.202005319},
    doi = {10.1002/adfm.202005319},
    issn = {16163028}
}

@article{Kadic2013MetamaterialsElectromagnetism,
    title = {{Metamaterials beyond electromagnetism}},
    year = {2013},
    journal = {Reports Progress in Physics},
    author = {Kadic, Muamer and B{\"{u}}ckmann, Tiemo and Schittny, Robert and Wegener, Martin},
    number = {12},
    month = {11},
    pages = {126501},
    volume = {76},
    publisher = {IOP Publishing},
    issn = {00344885}
}

@article{Coulais2018Multi-stepMetamaterials,
    title = {{Multi-step self-guided pathways for shape-changing metamaterials}},
    year = {2018},
    journal = {Nature},
    author = {Coulais, Corentin and Sabbadini, Alberico and Vink, Fré and van Hecke, Martin},
    number = {7724},
    month = {9},
    pages = {512},
    volume = {561},
    publisher = {Nature Publishing Group},
    url = {https://www.nature.com/articles/s41586-018-0541-0},
    doi = {10.1038/s41586-018-0541-0},
    issn = {14764687},
    pmid = {30258138},
    arxivId = {1810.07605}
}

@article{Shan2015MultistableEnergy,
    title = {{Multistable Architected Materials for Trapping Elastic Strain Energy}},
    year = {2015},
    journal = {Advanced Materials},
    author = {Shan, Sicong and Kang, Sung H. and Raney, Jordan R. and Wang, Pai and Fang, Lichen and Candido, Francisco and Lewis, Jennifer A. and Bertoldi, Katia},
    number = {29},
    month = {8},
    pages = {4296},
    volume = {27},
    publisher = {John Wiley {\&} Sons, Ltd},
    url = {https://onlinelibrary.wiley.com/doi/full/10.1002/adma.201501708 https://onlinelibrary.wiley.com/doi/abs/10.1002/adma.201501708 https://onlinelibrary.wiley.com/doi/10.1002/adma.201501708},
    doi = {10.1002/adma.201501708},
    issn = {15214095}
}

@article{Haghpanah2016MultistableMaterials,
    title = {{Multistable Shape-Reconfigurable Architected Materials}},
    year = {2016},
    journal = {Advanced Materials},
    author = {Haghpanah, Babak and Salari-Sharif, Ladan and Pourrajab, Peyman and Hopkins, Jonathan and Valdevit, Lorenzo},
    number = {36},
    month = {9},
    pages = {7915},
    volume = {28},
    publisher = {John Wiley {\&} Sons, Ltd},
    url = {https://onlinelibrary-wiley-com.proxy.library.upenn.edu/doi/full/10.1002/adma.201601650 https://onlinelibrary-wiley-com.proxy.library.upenn.edu/doi/abs/10.1002/adma.201601650 https://onlinelibrary-wiley-com.proxy.library.upenn.edu/doi/10.1002/adma.2016016},
    doi = {10.1002/adma.201601650},
    issn = {15214095}
}

@article{Chen2014NonlinearInsulator,
    title = {{Nonlinear conduction via solitons in a topological mechanical insulator}},
    year = {2014},
    journal = {Proceedings of the National Academy of Sciences of the United States of America},
    author = {Chen, Bryan Gin Ge and Upadhyaya, Nitin and Vitelli, Vincenzo},
    number = {36},
    pages = {13004},
    volume = {111},
    publisher = {National Academy of Sciences},
    url = {www.pnas.org/cgi/doi/10.1073/pnas.1405969111},
    doi = {10.1073/pnas.1405969111},
    issn = {10916490},
    pmid = {25157161},
    arxivId = {1404.2263}
}

@article{Guest2003OnStructures,
    title = {{On the determinacy of repetitive structures}},
    year = {2003},
    journal = {Journal of the Mechanics and Physics of Solids},
    author = {Guest, S. D. and Hutchinson, J. W.},
    number = {3},
    month = {3},
    pages = {383},
    volume = {51},
    publisher = {Pergamon},
    doi = {10.1016/S0022-5096(02)00107-2},
    issn = {00225096}
}

@article{Kadic2012OnMetamaterials,
    title = {{On the practicability of pentamode mechanical metamaterials}},
    year = {2012},
    journal = {Applied Physics Letters},
    author = {Kadic, Muamer and B{\"{u}}ckmann, Tiemo and Stenger, Nicolas and Thiel, Michael and Wegener, Martin},
    number = {19},
    month = {5},
    pages = {191901},
    volume = {100},
    publisher = {American Institute of PhysicsAIP},
    url = {https://aip.scitation.org/doi/abs/10.1063/1.4709436},
    doi = {10.1063/1.4709436},
    issn = {00036951}
}

@article{Florijn2014ProgrammableMetamaterials,
    title = {{Programmable mechanical metamaterials}},
    year = {2014},
    journal = {Physical Review Letters},
    author = {Florijn, Bastiaan and Coulais, Corentin and Van Hecke, Martin},
    number = {17},
    month = {10},
    pages = {175503},
    volume = {113},
    publisher = {American Physical Society},
    url = {https://journals.aps.org/prl/abstract/10.1103/PhysRevLett.113.175503},
    doi = {10.1103/PhysRevLett.113.175503},
    issn = {10797114},
    arxivId = {1407.4273}
}

@article{Paulose2015SelectiveMetamaterials,
    title = {{Selective buckling via states of self-stress in topological metamaterials}},
    year = {2015},
    journal = {Proceedings of the National Academy of Sciences of the United States of America},
    author = {Paulose, Jayson and Meeussen, Anne S. and Vitelli, Vincenzo},
    number = {25},
    month = {6},
    pages = {7639},
    volume = {112},
    publisher = {National Academy of Sciences},
    doi = {10.1073/pnas.1502939112},
    issn = {10916490},
    arxivId = {1502.03396}
}

@article{Farzaneh2022SequentialRatios,
    title = {{Sequential metamaterials with alternating Poisson’s ratios}},
    year = {2022},
    journal = {Nature Communications},
    author = {Farzaneh, Amin and Pawar, Nikhil and Portela, Carlos M. and Hopkins, Jonathan B.},
    number = {1},
    month = {2},
    pages = {1},
    volume = {13},
    publisher = {Nature Publishing Group},
    url = {https://www.nature.com/articles/s41467-022-28696-9},
    doi = {10.1038/s41467-022-28696-9},
    issn = {20411723},
    pmid = {35210416}
}

@article{Stenull2019SignaturesLattices,
    title = {{Signatures of Topological Phonons in Superisostatic Lattices}},
    year = {2019},
    journal = {Physical Review Letters},
    author = {Stenull, Olaf and Lubensky, T. C.},
    number = {24},
    month = {6},
    volume = {122},
    publisher = {American Physical Society},
    issn = {10797114},
    pmid = {31322362},
    arxivId = {1906.10601}
}

@article{Zunker2021SoftWheels,
    title = {{Soft topological lattice wheels}},
    year = {2021},
    journal = {Extreme Mechanics Letters},
    author = {Zunker, William and Gonella, Stefano},
    month = {7},
    pages = {101344},
    volume = {46},
    publisher = {Elsevier},
    doi = {10.1016/j.eml.2021.101344},
    issn = {23524316},
    arxivId = {2012.12832}
}

@article{Janbaz2020StrainMetamaterials,
    title = {{Strain rate-dependent mechanical metamaterials}},
    year = {2020},
    journal = {Science Advances},
    author = {Janbaz, S. and Narooei, K. and Van Manen, T. and Zadpoor, A. A.},
    number = {25},
    month = {6},
    pages = {616},
    volume = {6},
    publisher = {American Association for the Advancement of Science},
    url = {https://www.science.org/doi/full/10.1126/sciadv.aba0616},
    doi = {10.1126/sciadv.aba0616},
    issn = {23752548},
    pmid = {32596451}
}

@article{Frenzel2016TailoredAbsorbers,
    title = {{Tailored Buckling Microlattices as Reusable Light-Weight Shock Absorbers}},
    year = {2016},
    journal = {Advanced Materials},
    author = {Frenzel, Tobias and Findeisen, Claudio and Kadic, Muamer and Gumbsch, Peter and Wegener, Martin},
    number = {28},
    month = {7},
    pages = {5865},
    volume = {28},
    publisher = {John Wiley {\&} Sons, Ltd},
    url = {https://onlinelibrary-wiley-com.proxy.library.upenn.edu/doi/full/10.1002/adma.201600610 https://onlinelibrary-wiley-com.proxy.library.upenn.edu/doi/abs/10.1002/adma.201600610 https://onlinelibrary-wiley-com.proxy.library.upenn.edu/doi/10.1002/adma.2016006},
    doi = {10.1002/adma.201600610},
    issn = {15214095},
    pmid = {27159205}
}

@article{Zhang2021TailoredIsolation,
    title = {{Tailored Mechanical Metamaterials with Programmable Quasi-Zero-Stiffness Features for Full-Band Vibration Isolation}},
    year = {2021},
    journal = {Advanced Functional Materials},
    author = {Zhang, Quan and Guo, Dengke and Hu, Gengkai},
    number = {33},
    month = {8},
    pages = {2101428},
    volume = {31},
    publisher = {John Wiley {\&} Sons, Ltd},
    url = {https://onlinelibrary.wiley.com/doi/full/10.1002/adfm.202101428 https://onlinelibrary.wiley.com/doi/abs/10.1002/adfm.202101428 https://onlinelibrary.wiley.com/doi/10.1002/adfm.202101428},
    doi = {10.1002/adfm.202101428},
    issn = {16163028}
}

@article{Frenzel2017Three-dimensionalTwist,
    title = {{Three-dimensional mechanical metamaterials with a twist}},
    year = {2017},
    journal = {Science},
    author = {Frenzel, Tobias and Kadic, Muamer and Wegener, Martin},
    number = {6366},
    month = {11},
    pages = {1072},
    volume = {358},
    publisher = {American Association for the Advancement of Science},
    url = {https://www.science.org},
    doi = {10.1126/science.aao4640},
    issn = {10959203},
    pmid = {29170236}
}

@article{Chen2016TopologicalKirigami,
    title = {{Topological Mechanics of Origami and Kirigami}},
    year = {2016},
    journal = {Physical Review Letters},
    author = {Chen, Bryan Gin Ge and Liu, Bin and Evans, Arthur A. and Paulose, Jayson and Cohen, Itai and Vitelli, Vincenzo and Santangelo, C. D.},
    number = {13},
    month = {3},
    pages = {135501},
    volume = {116},
    publisher = {American Physical Society},
    url = {https://journals-aps-org.proxy.library.upenn.edu/prl/abstract/10.1103/PhysRevLett.116.135501},
    doi = {10.1103/PhysRevLett.116.135501},
    issn = {10797114},
    pmid = {27081987},
    arxivId = {1508.00795}
}

@article{Paulose2015TopologicalMetamaterials,
    title = {{Topological modes bound to dislocations in mechanical metamaterials}},
    year = {2015},
    journal = {Nature Physics},
    author = {Paulose, Jayson and Chen, Bryan Gin Ge and Vitelli, Vincenzo},
    number = {2},
    month = {1},
    pages = {153},
    volume = {11},
    publisher = {Nature Publishing Group},
    url = {https://www-nature-com.proxy.library.upenn.edu/articles/nphys3185},
    doi = {10.1038/nphys3185},
    issn = {17452481},
    arxivId = {1406.3323}
}

@article{Stenull2016TopologicalDimensions,
    title = {{Topological Phonons and Weyl Lines in Three Dimensions}},
    year = {2016},
    journal = {Physical Review Letters},
    author = {Stenull, Olaf and Kane, C. L. and Lubensky, T. C.},
    number = {6},
    month = {8},
    pages = {068001},
    volume = {117},
    publisher = {American Physical Society},
    url = {https://journals-aps-org.proxy.library.upenn.edu/prl/abstract/10.1103/PhysRevLett.117.068001},
    issn = {10797114},
    arxivId = {1606.00301}
}

@article{Hwang2018TunableStructures,
    title = {{Tunable Mechanical Metamaterials through Hybrid Kirigami Structures}},
    year = {2018},
    journal = {Scientific Reports},
    author = {Hwang, Doh Gyu and Bartlett, Michael D.},
    number = {1},
    month = {2},
    pages = {1},
    volume = {8},
    publisher = {Nature Publishing Group},
    url = {https://www.nature.com/articles/s41598-018-21479-7},
    doi = {10.1038/s41598-018-21479-7},
    issn = {20452322},
    pmid = {29467413}
}

@article{Zheng2014UltralightMetamaterials,
    title = {{Ultralight, ultrastiff mechanical metamaterials}},
    year = {2014},
    journal = {Science},
    author = {Zheng, Xiaoyu and Lee, Howon and Weisgraber, Todd H. and Shusteff, Maxim and DeOtte, Joshua and Duoss, Eric B. and Kuntz, Joshua D. and Biener, Monika M. and Ge, Qi and Jackson, Julie A. and Kucheyev, Sergei O. and Fang, Nicholas X. and Spadaccini, Christopher M.},
    number = {6190},
    month = {6},
    pages = {1373},
    volume = {344},
    publisher = {American Association for the Advancement of Science},
    url = {https://www.science.org/doi/full/10.1126/science.1252291},
    doi = {10.1126/science.1252291},
    issn = {10959203},
    pmid = {24948733}
}
\end{document}


\title{\begin{center}
    Supporting Information \newline \end{center} Soft mechanical metamaterials with transformable topology protected by stress caching
}
\author{Jason~Christopher Jolly\textsuperscript{1}, Binjie~Jin\textsuperscript{2}, Lishuai~Jin\textsuperscript{1}, YoungJoo~Lee\textsuperscript{1}, Tao~Xie\textsuperscript{2}, Stefano~Gonella\textsuperscript{3}, Kai~Sun\textsuperscript{4}, Xiaoming~Mao\textsuperscript{4}\thanks{Corresponding author.}\thanksspace \space and Shu~Yang\textsuperscript{1}\footnotemark[1]\thanksspace ~\\[-3pt]\normalsize\normalfont ~\\
\textsuperscript{1}{Department of Materials Science and Engineering\unskip, University of Pennsylvania\unskip, 3231 Walnut Street\unskip, Philadelphia\unskip, 19103\unskip, Pennsylvania\unskip, USA}~\\
\textsuperscript{2}{State Key Laboratory of Chemical Engineering, Department of Chemical and Biological Engineering\unskip, Zhejiang University\unskip, 38 Zhe Da Road\unskip, Hangzhou\unskip, 310027\unskip, Zhejiang\unskip, China}~\\
\textsuperscript{3}{Department of Civil, Environmental, and Geo- Engineering\unskip, University of Minnesota\unskip, 500 Pillsbury Drive S.E.\unskip, Minneapolis\unskip, 55455\unskip, Minnesota\unskip, USA}~\\
\textsuperscript{4}{Department of Physics\unskip, University of Michigan\unskip, 450 Church St.\unskip, Ann Arbor\unskip, 48109\unskip, Michigan\unskip, USA\newline Corresponding E-mail: maox@umich.edu (Xiaoming~Mao), shuyang@seas.upenn.edu (Shu~Yang)}}

\def\RunningHead{}\def\RunningAuthor{Christopher Jolly \MakeLowercase{\textit{et al.}} }

\maketitle 

\setcounter{figure}{0}
\renewcommand{\thefigure}{S\arabic{figure}}

\begin{figure}[!htp]
\begin{center}
\includegraphics[width=1\textwidth]{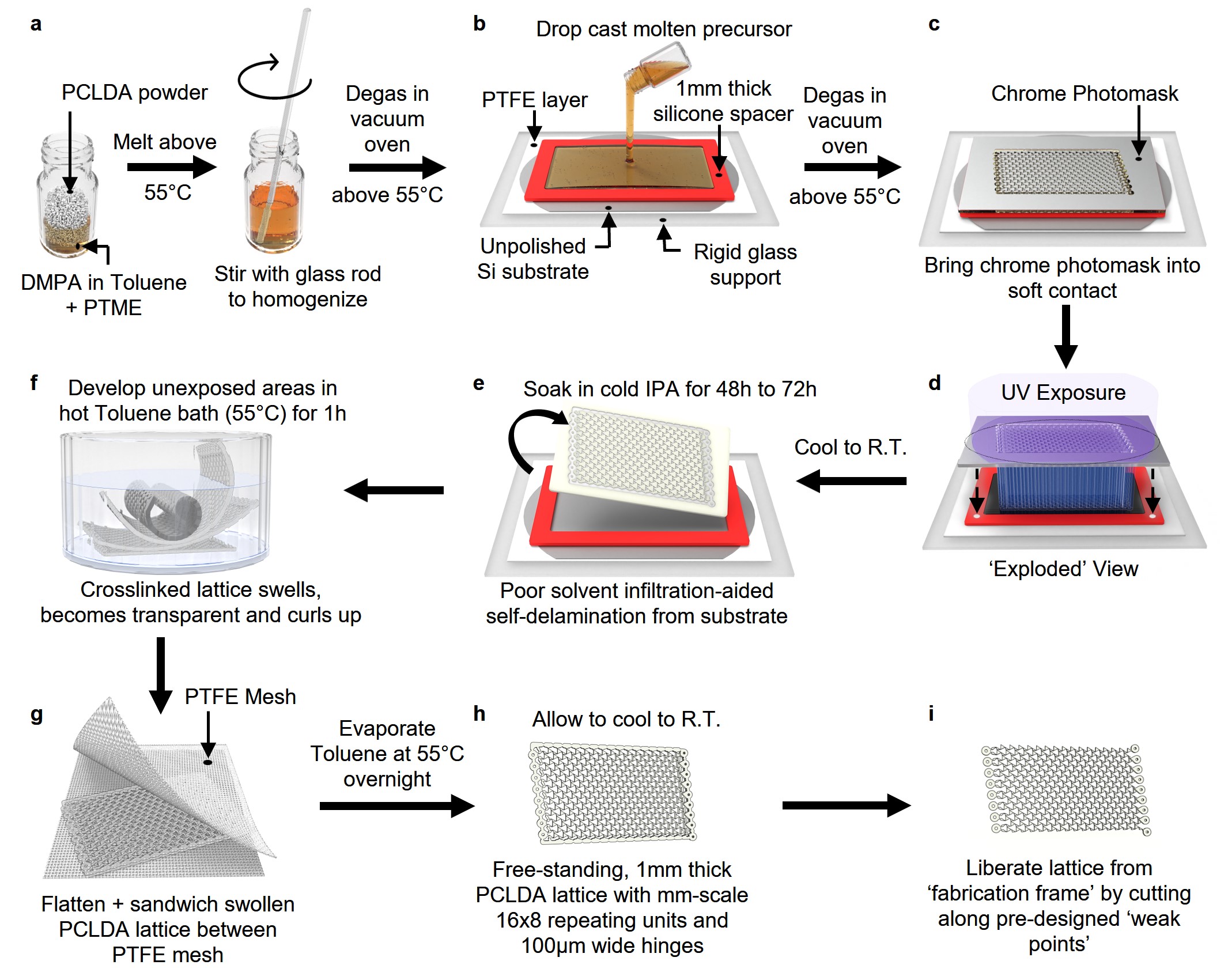}
\caption{Fabrication of a free-standing PCLDA-SMP TTMM lattice via a multi-step ‘thick’ photolithography process.}
\label{Figure_S1}
\end{center}
\end{figure}

\begin{figure}[!htp]
\begin{center}
\includegraphics[width=1\textwidth]{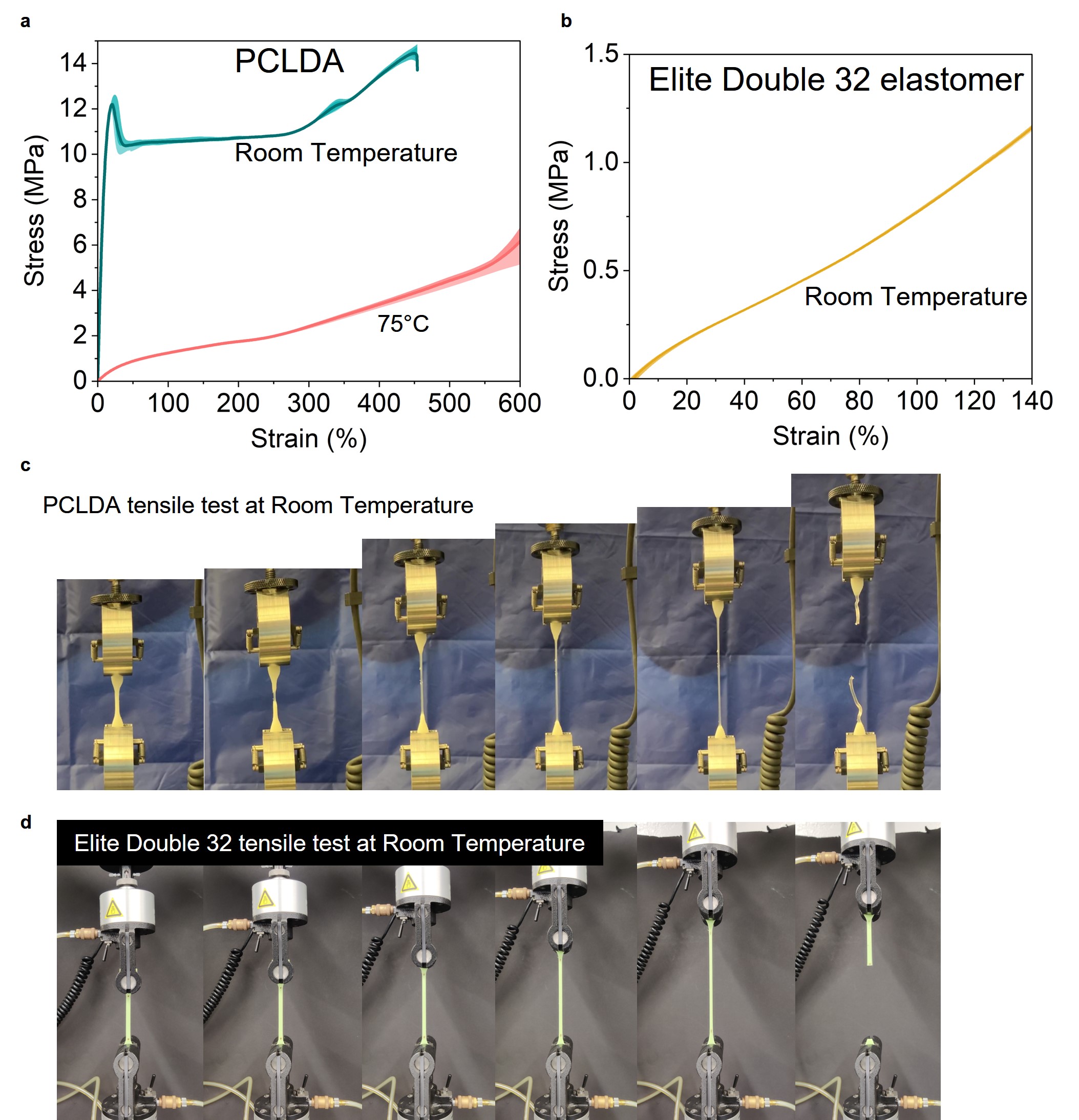}
\caption{{a,c) Tensile test results of PCLDA-SMP dog bone samples both at room temperature and above its melting temperature and b,d) Tensile test results of Elite Double 32 dog bone samples at room temperature.}}
\label{Figure_S2}
\end{center}
\end{figure}

\begin{figure}[!htp]
\begin{center}
\includegraphics[width=1\textwidth]{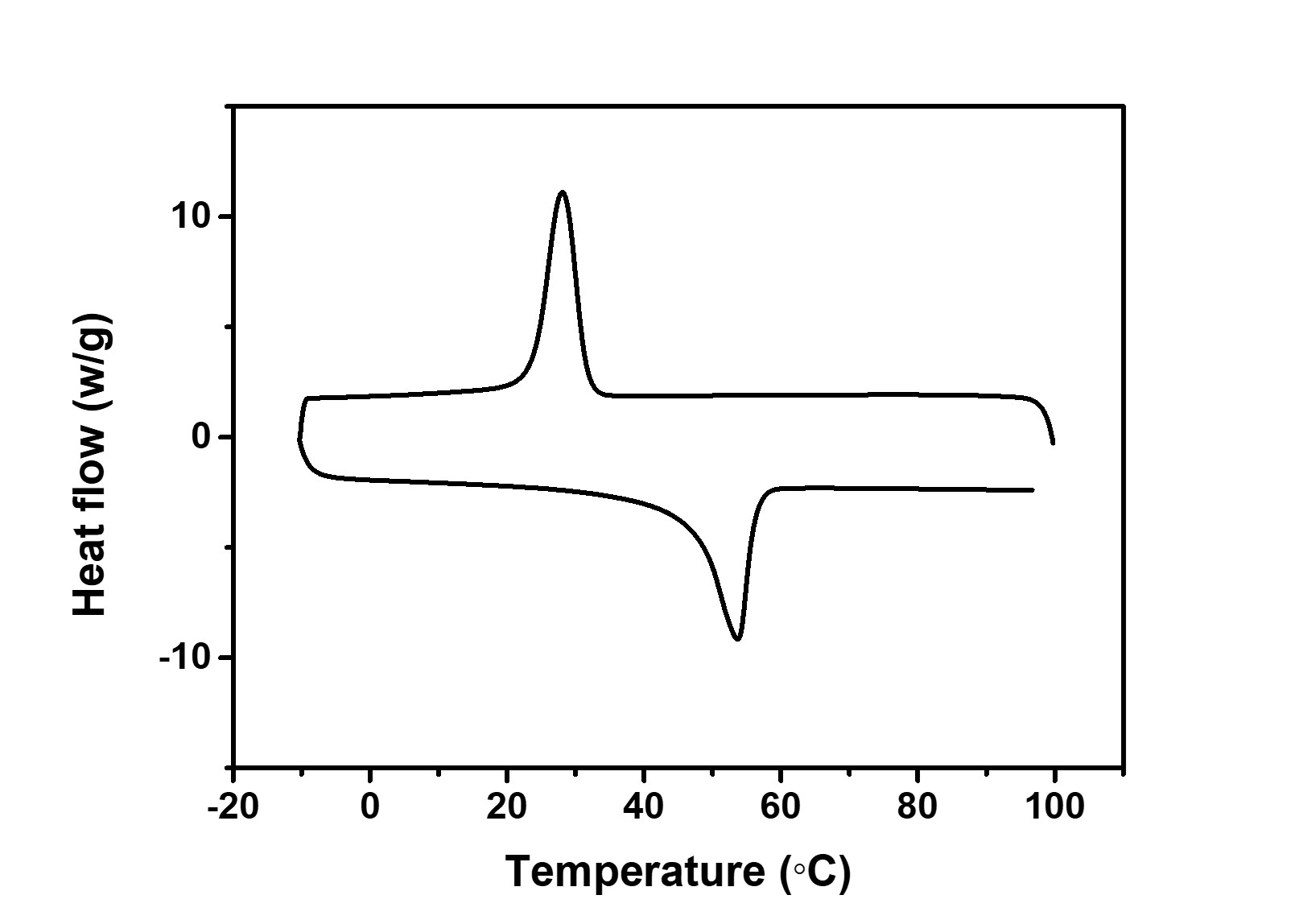}
\caption{{Characterization of the melting and crystallization transition temperatures of PCLDA-SMP via Differential Scanning Calorimetry (DSC).}}
\label{Figure_S3}
\end{center}
\end{figure}

\begin{figure}[!htp]
\begin{center}
\includegraphics[width=1\textwidth]{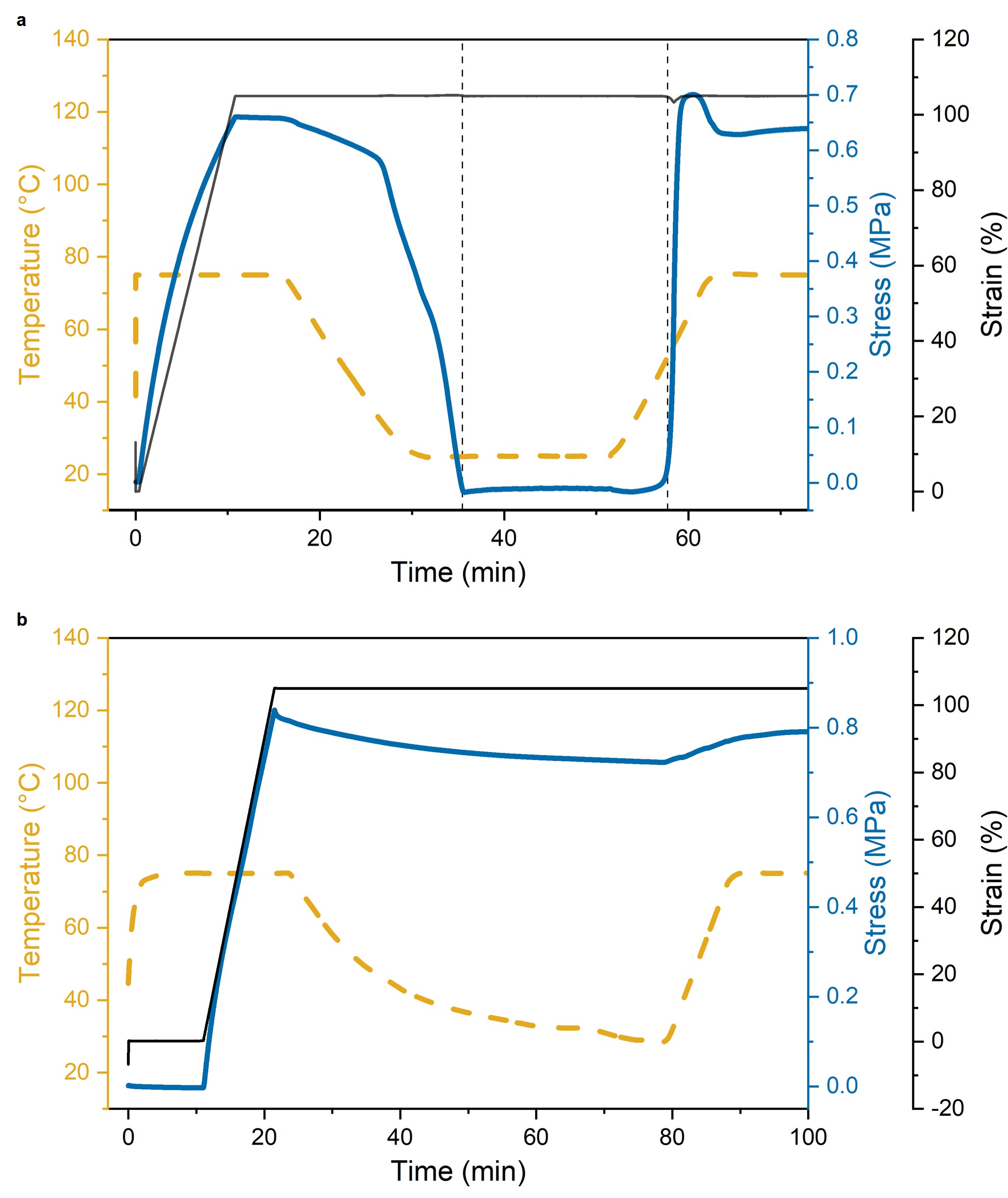}
\caption{{a) Characterizing the stress caching ability of PCLDA-SMP via Dynamic Mechanical Analysis (DMA) i.e., its ability to ‘lock away’ any stresses generated upon stretching (to a maximum strain of 105\%), upon cooling below its melting temperature and to maintain a ‘zero residual stress state’ while at room temperature. These restorative elastic stress are unlocked upon reheating above the melting temperature. b) A similar DMA test performed on an Elite Double 32 elastomer sample which serves as a reference case for a polymer system without stress caching, wherein nearly all generated stress remains ‘unlocked’, always.}}
\label{Figure_S4}
\end{center}
\end{figure}

\begin{figure}[!htp]
\begin{center}
\includegraphics[width=1\textwidth]{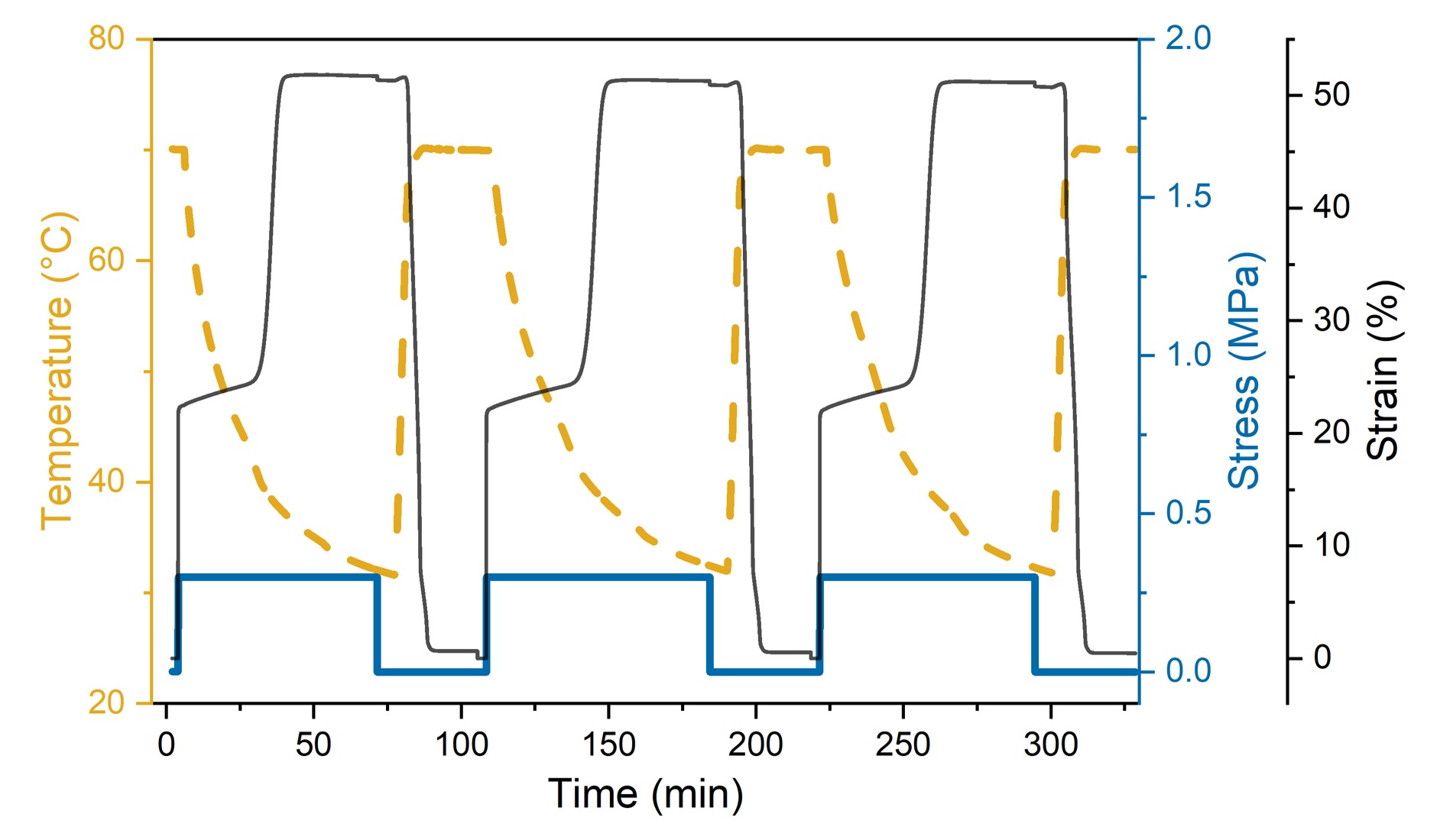}
\caption{{Characterizing the robustness and repeatability of the intrinsic shape memory mechanism of PCLDA-SMP via Dynamic Mechanical Analysis (DMA) over 3x cycles.}}
\label{Figure_S5}
\end{center}
\end{figure}

\begin{figure}[!htp]
\begin{center}
\includegraphics[width=1\textwidth]{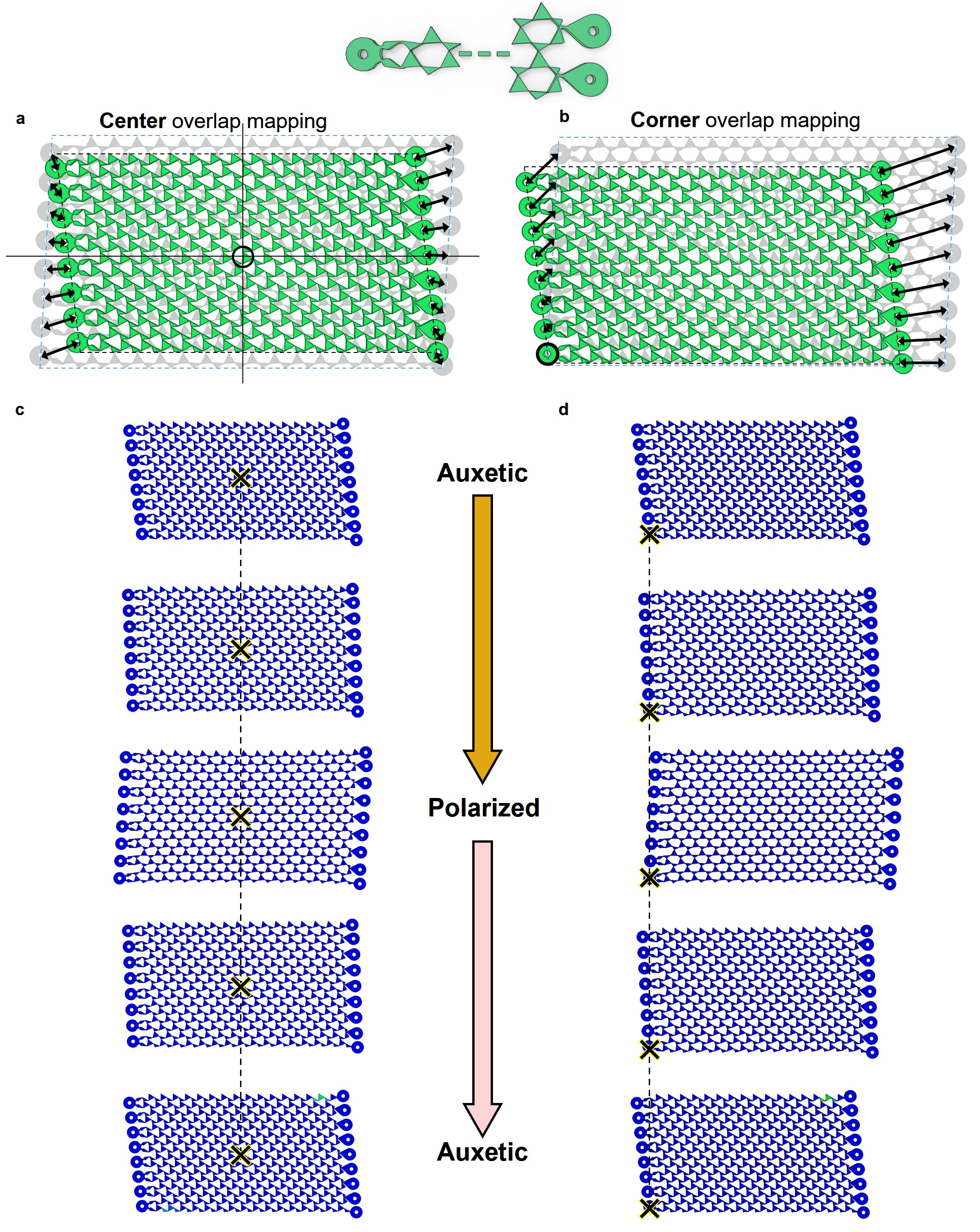}
\caption{{Mapping the edges of initial and final lattice configurations determines the vector displacements required to prescribe kinematic transformations. a,b) Overlaying auxetic and polarized variants about (a) their centers and (b) one of their corners, permits the definition of vectors that map the corners of triangular units or centers of edge loops along ‘input’ edges to their equivalent positions in a transformed lattice. c,d) Snapshots of simulation results that confirm the equivalence of kinematic transformations with vectors defined using different common overlay points, in prescribing a reversible transformation of the same lattice between the same two auxetic and polarized states.}}
\label{Figure_S6}
\end{center}
\end{figure}

\begin{figure}[!htp]
\begin{center}
\includegraphics[width=1\textwidth]{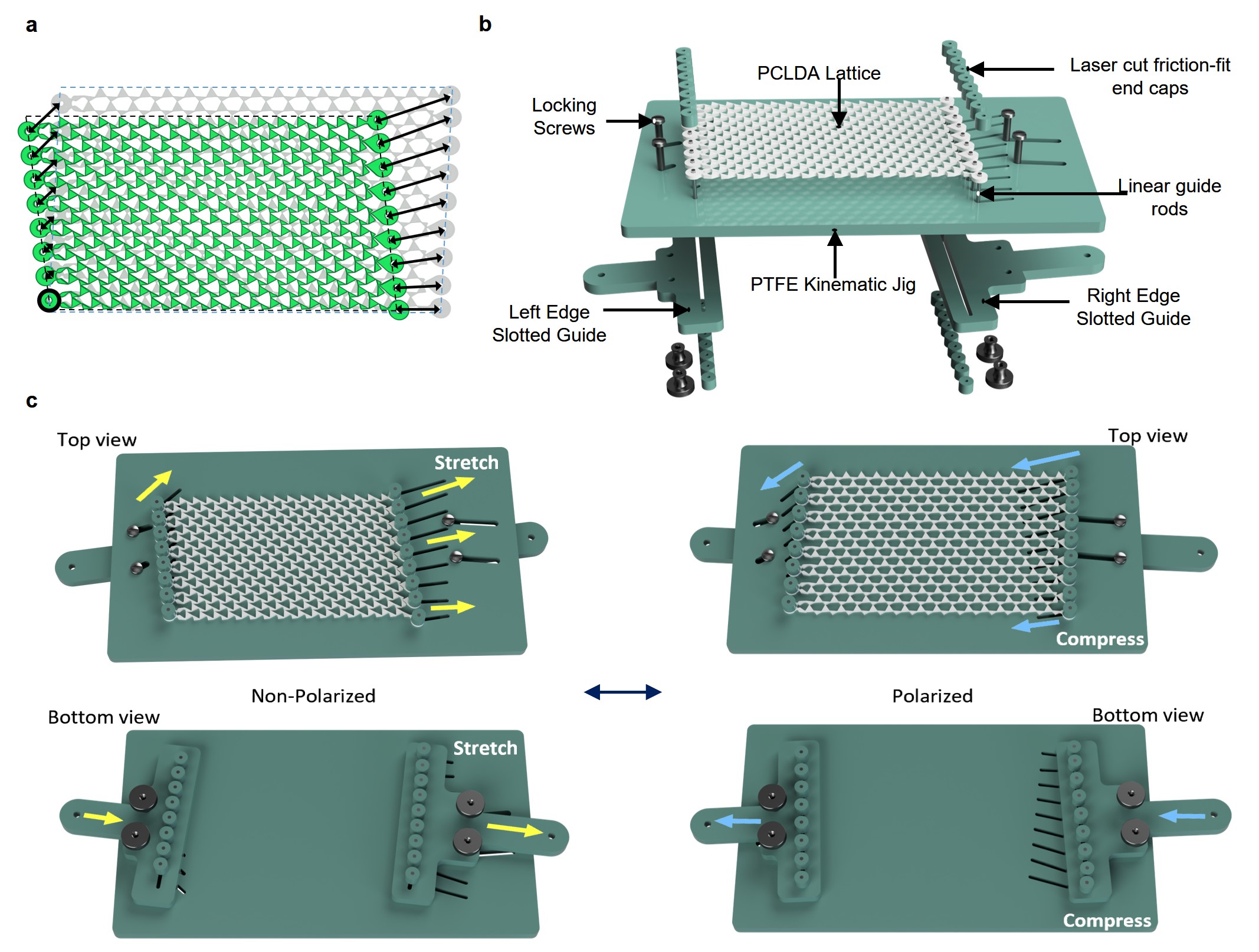}
\caption{{Design and assembly of a K-2 PCLDA-SMP lattice and its kinematic jig. a) The vector inputs and by extension, the design of the slots to be cut into the kinematic jig, are obtained by overlaying auxetic and polarized lattice conformations about a shared corner. b) Assembling the various components of the kinematic jig and the PCLDA-SMP lattice. c) Schematic of the lattice being transformed between its auxetic and polarized phases upon vectored stretching and compression in the physical jig, respectively.}}
\label{Figure_S7}
\end{center}
\end{figure}

\begin{figure}[!htp]
\begin{center}
\includegraphics[width=1\textwidth]{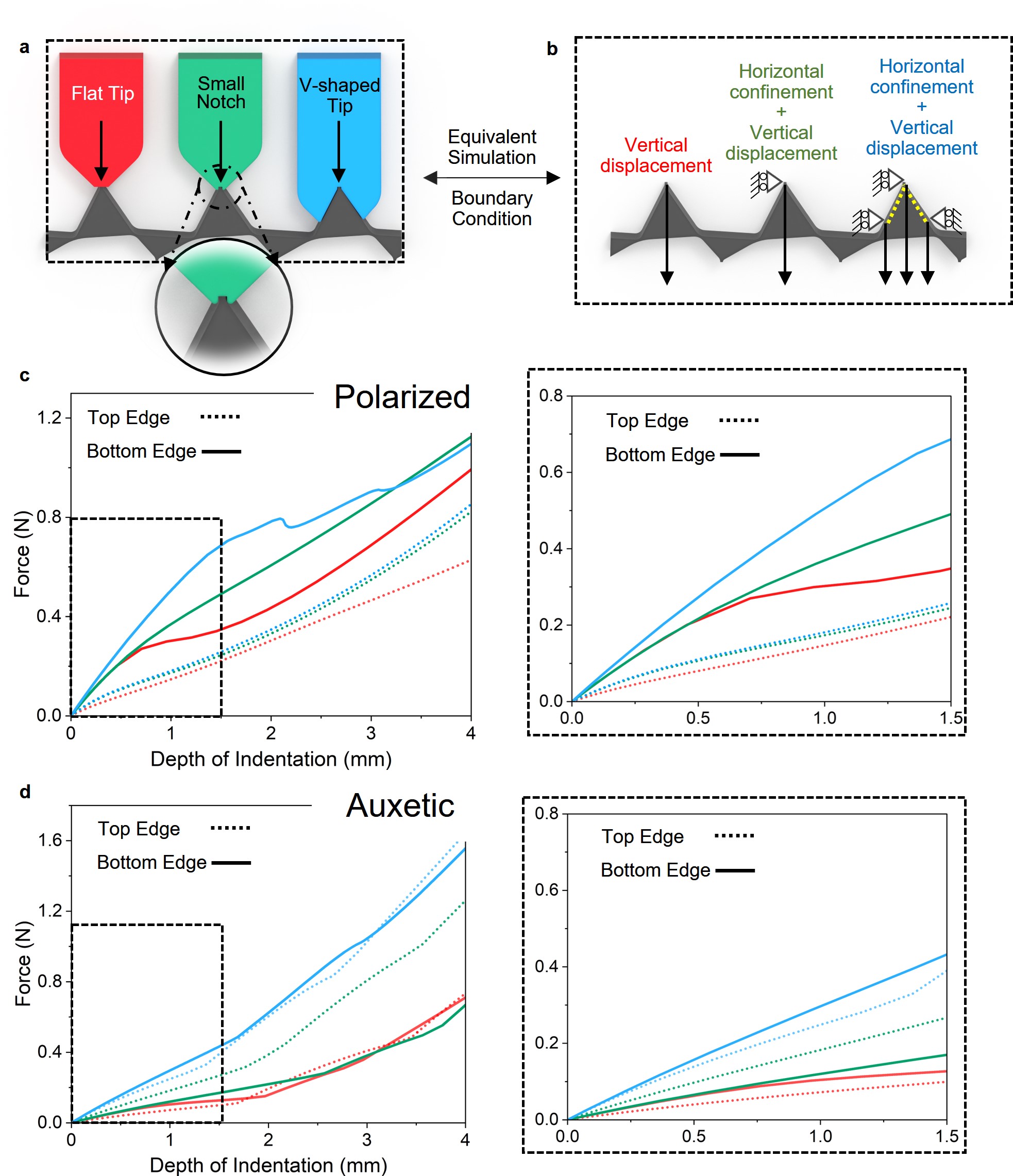}
\caption{{Effect of indenter tip geometry on topological edge behaviour. a) Three different indenter tip geometries and their influence on mechanical edge behavior were studied via FEM. b) Their respective interactions with the lattice were simulated as different boundary conditions at the tip and edges of the local triangular unit being indented. c,d) Force-displacement curves obtained by simulating the indentation of (c) polarized and (d) auxetic lattices with the aforementioned indenter tip geometries.}}
\label{Figure_S8}
\end{center}
\end{figure}

\begin{figure}[!htp]
\begin{center}
\includegraphics[width=1\textwidth]{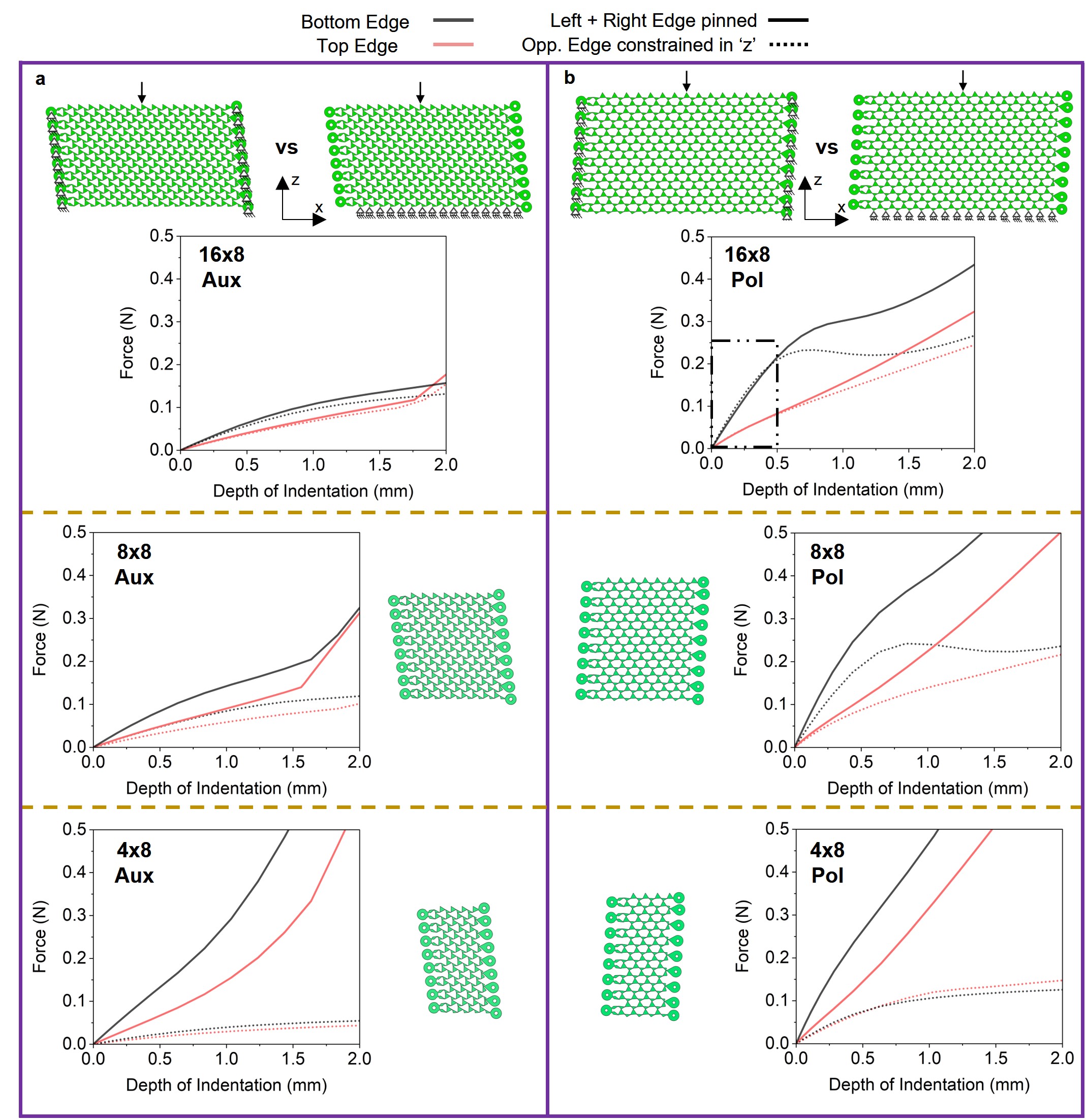}
\caption{{a,b) Effect of sample boundary conditions during indentation i.e., L-R edge pinning versus bottom edge pinning, on the topological behavior of three different sizes of (a) auxetic lattices and (b) polarized lattices.}}
\label{Figure_S9}
\end{center}
\end{figure}

\begin{figure}[!htp]
\begin{center}
\includegraphics[width=1\textwidth]{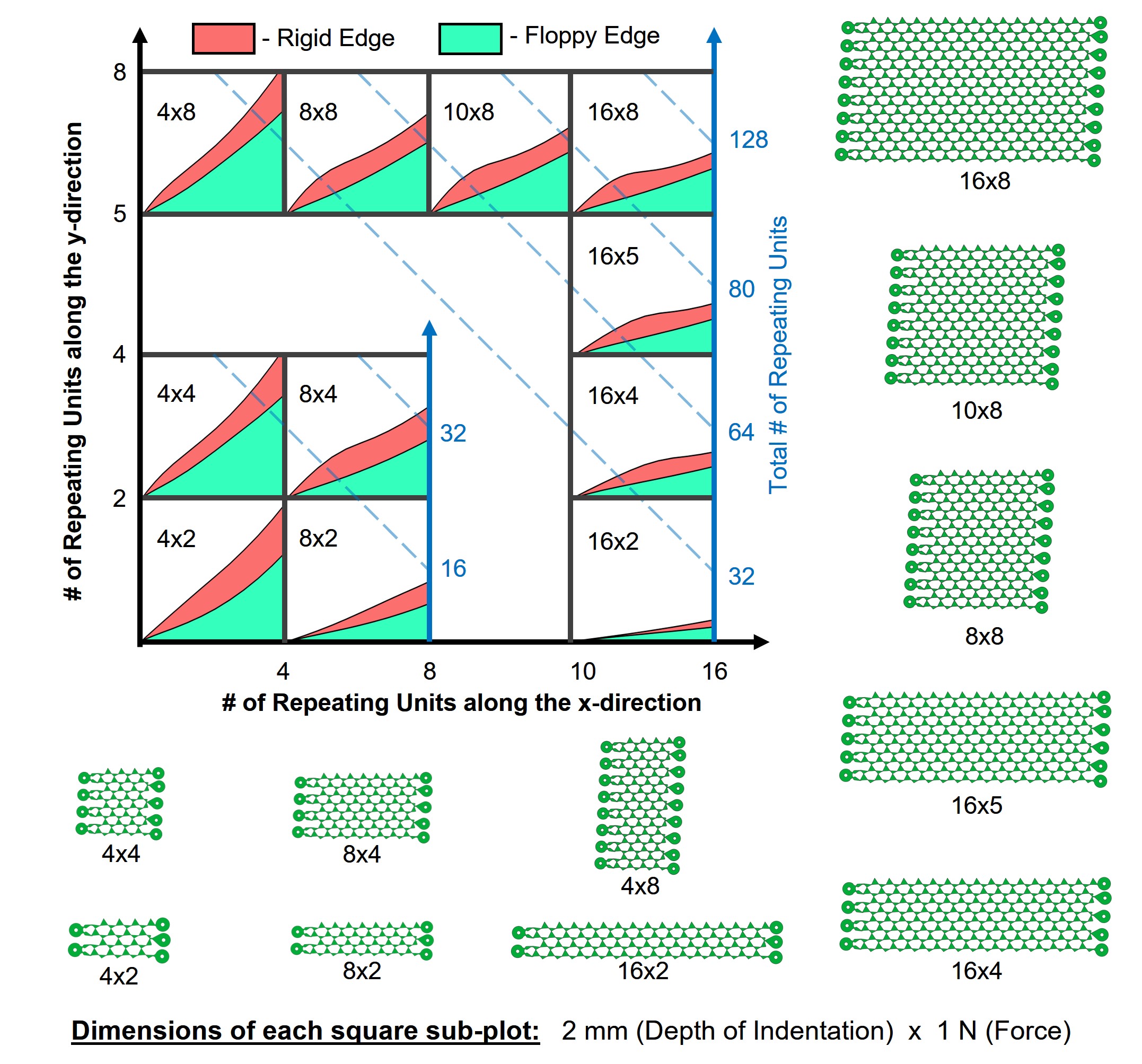}
\caption{{Effect of lattice size and aspect ratio on topological edge behavior.}}
\label{Figure_S10}
\end{center}
\end{figure}

\begin{figure}[!htp]
\begin{center}
\includegraphics[width=1\textwidth]{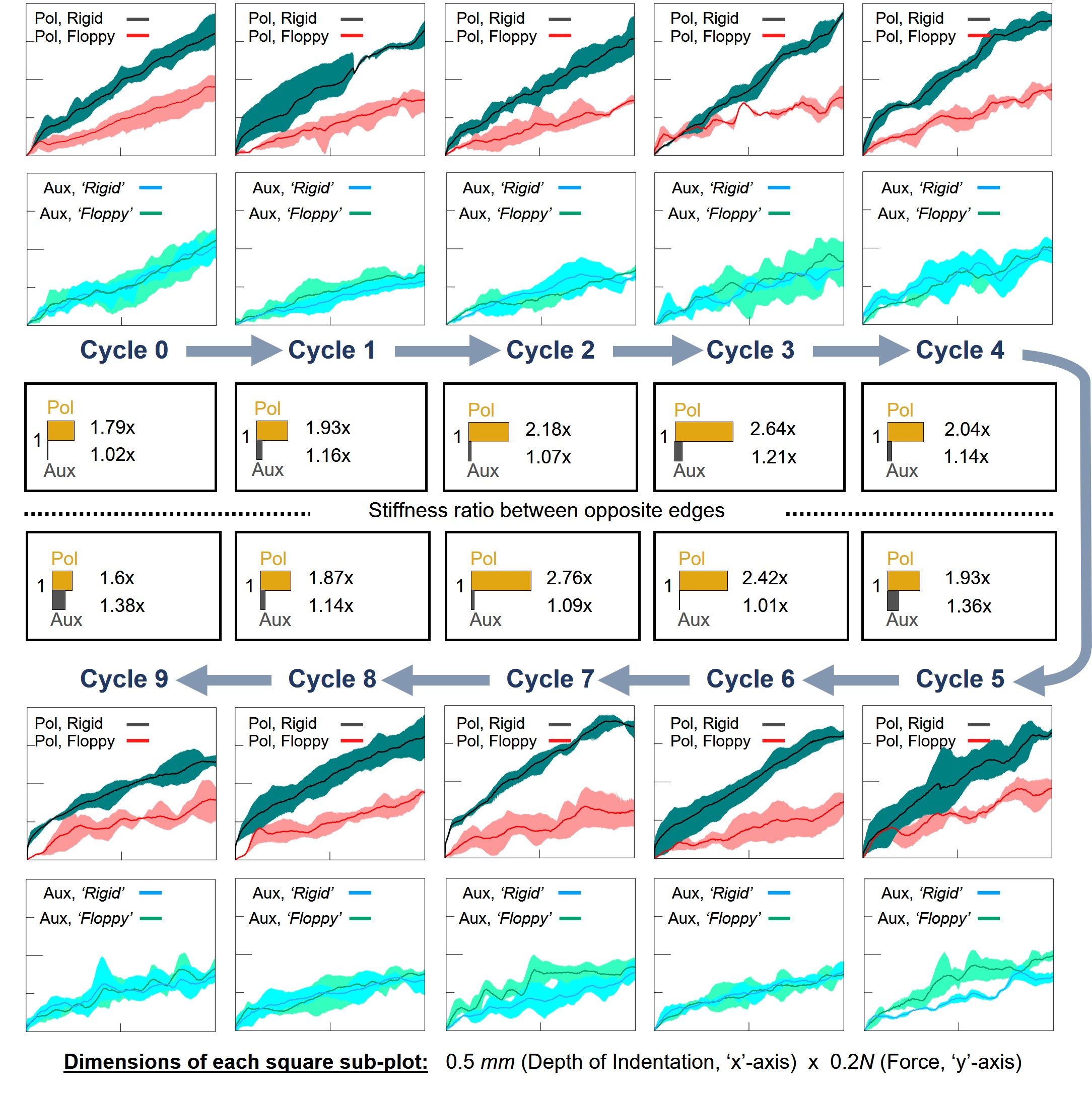}
\caption{{Edge indentation tests on auxetic and polarized configurations of the same PCLDA-SMP TTMM Lattice subjected to 10$\times$ complete kinematic cycles. Stiffness ratios (see histograms) were calculated from the slopes of the linear, small indentation regimes. The results reveal good cyclability between auxetic (S.R.$\sim$1.15) and polarized phases (S.R.$\sim$2.1)}}
\label{Figure_S11}
\end{center}
\end{figure}

\begin{figure}[!htp]
\begin{center}
\includegraphics[width=1\textwidth]{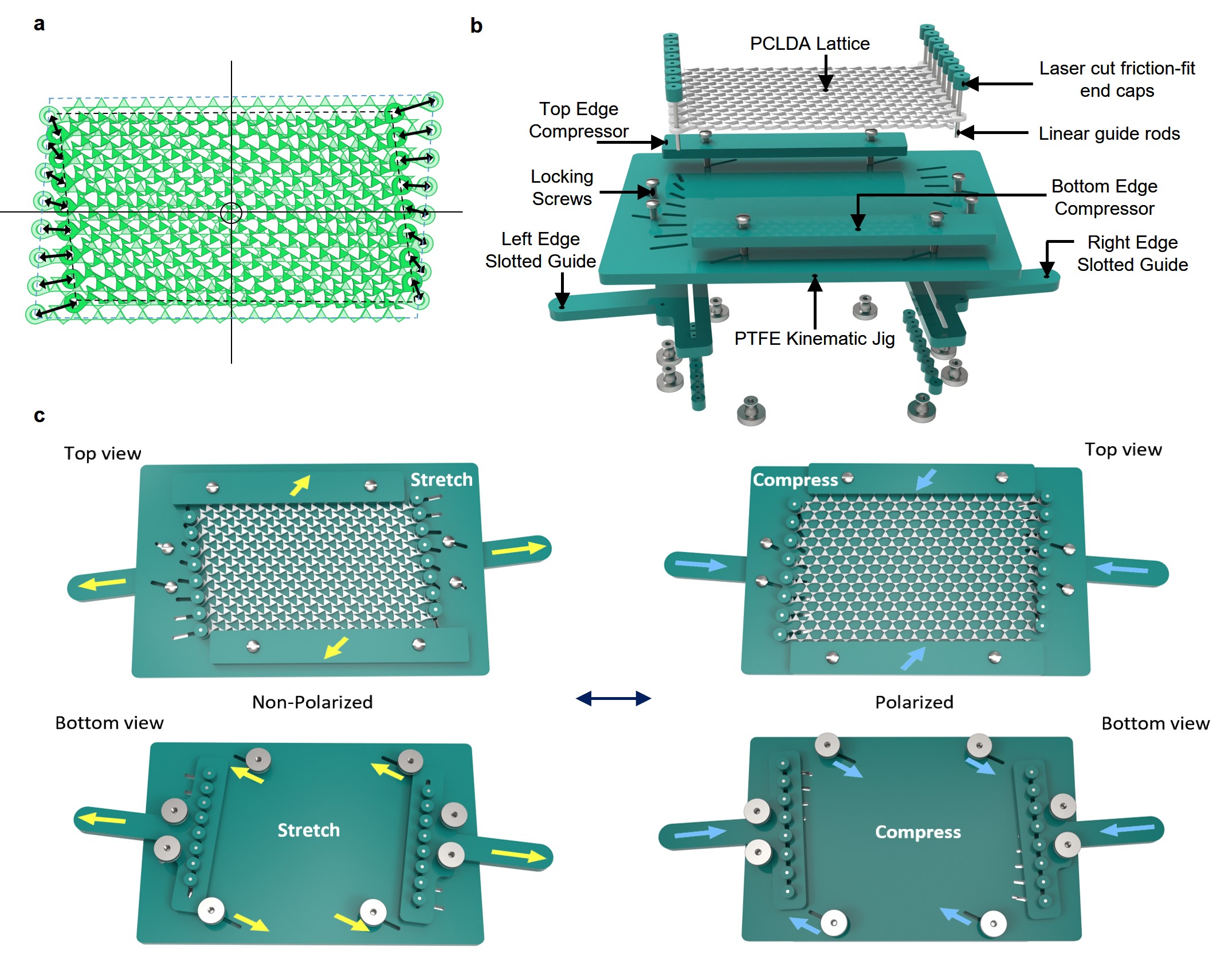}
\caption{{Design and assembly of a K-1 PCLDA-SMP lattice and its kinematic jig. a) The vector inputs and by extension, the design of the slots to be cut into the kinematic jig, are obtained by overlaying auxetic and polarized lattice conformations about a shared geometric center. b) Assembling the various components of the kinematic jig and the K-1 lattice. c) Schematic of the lattice being transformed between its auxetic and polarized phases upon vectored stretching and compression in the physical jig, respectively.}}
\label{Figure_S12}
\end{center}
\end{figure}

\begin{figure}[!htp]
\begin{center}
\includegraphics[width=1\textwidth]{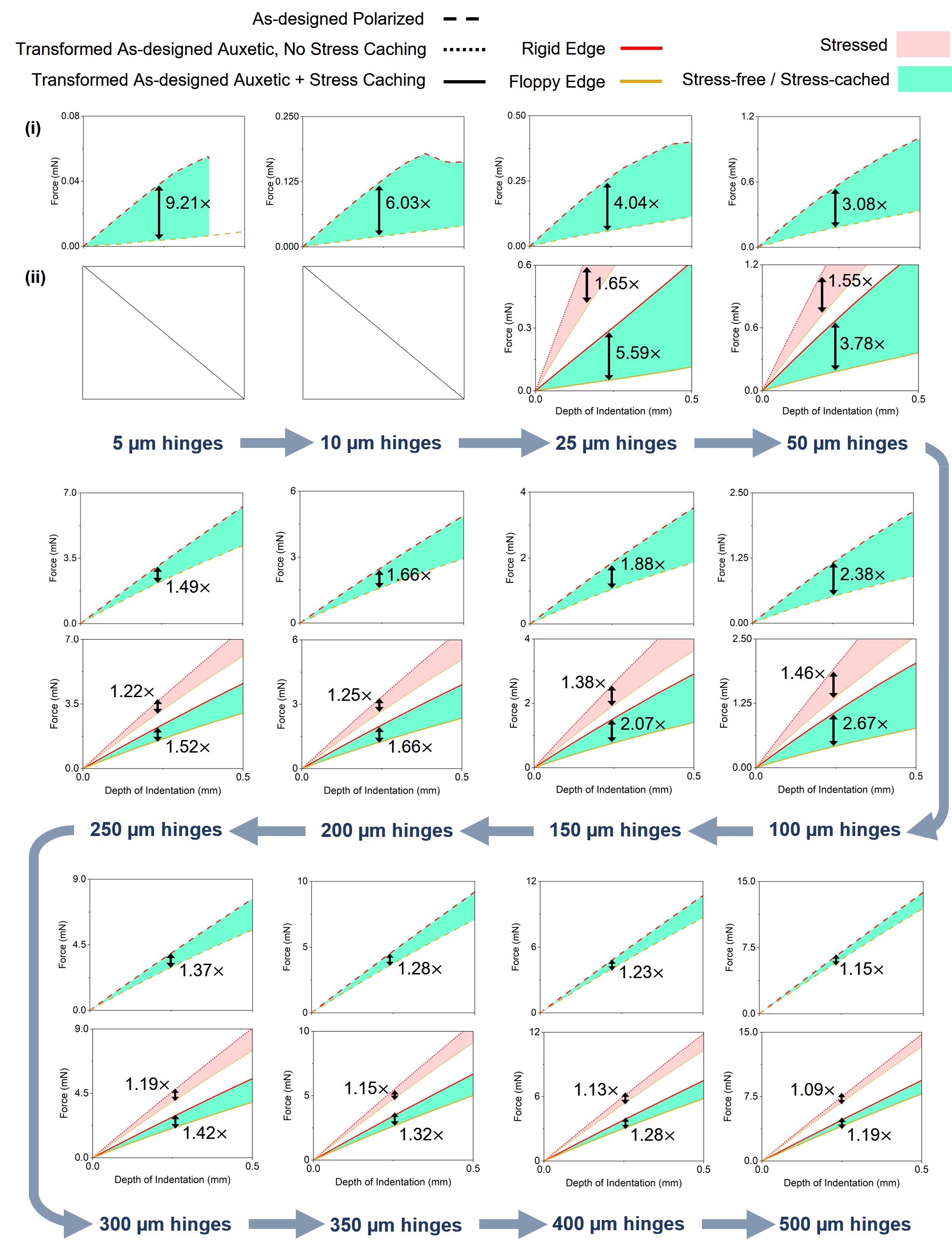}
\caption{{Simulated indentations on lattices with systematically varying hinge width and with material properties representative of the commercial elastomer (ED-32). i, ii) Force-displacement curves and stiffness ratios of (i) as-designed polarized lattices and (ii) as-designed auxetic structures that have been stretched into their polarized configurations, before and after stress caching. Simulations utilized the Gent hyperelastic model to capture the behavior of ED-32.}}
\label{Figure_S13}
\end{center}
\end{figure}

\begin{figure}[!htp]
\begin{center}
\includegraphics[width=1\textwidth]{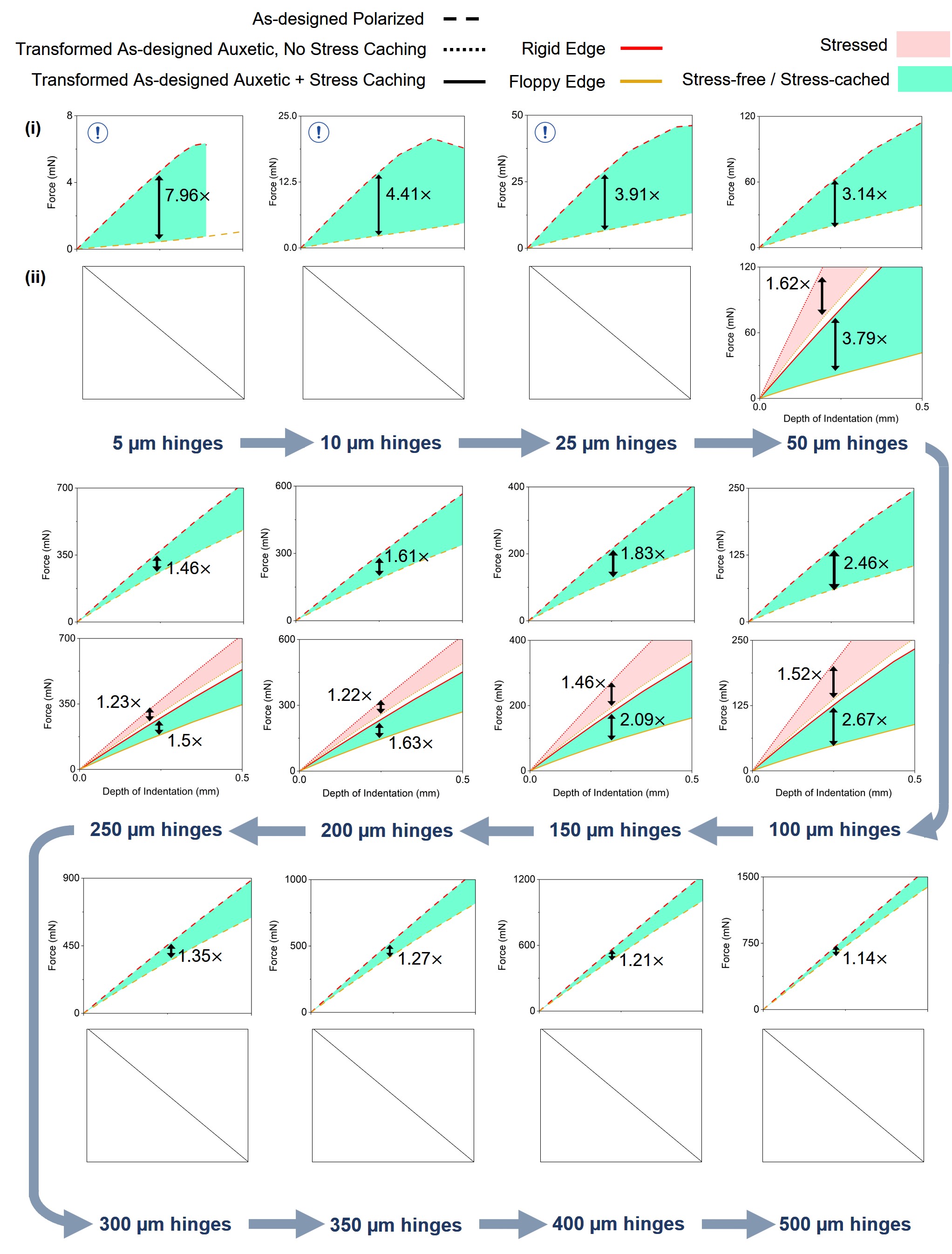}
\caption{{Simulated indentations on lattices with systematically varying hinge width and with material properties representative of PCLDA-SMP. i, ii) Force-displacement curves and stiffness ratios of (i) as-designed polarized lattices and (ii) as-designed auxetic structures that have been stretched into their polarized configurations, before and after stress caching. Simulations utilized an elastic-plastic model to capture the behavior of PCLDA-SMP.}}
\label{Figure_S14}
\end{center}
\end{figure}

\begin{figure}[!htp]
\begin{center}
\includegraphics[width=1\textwidth]{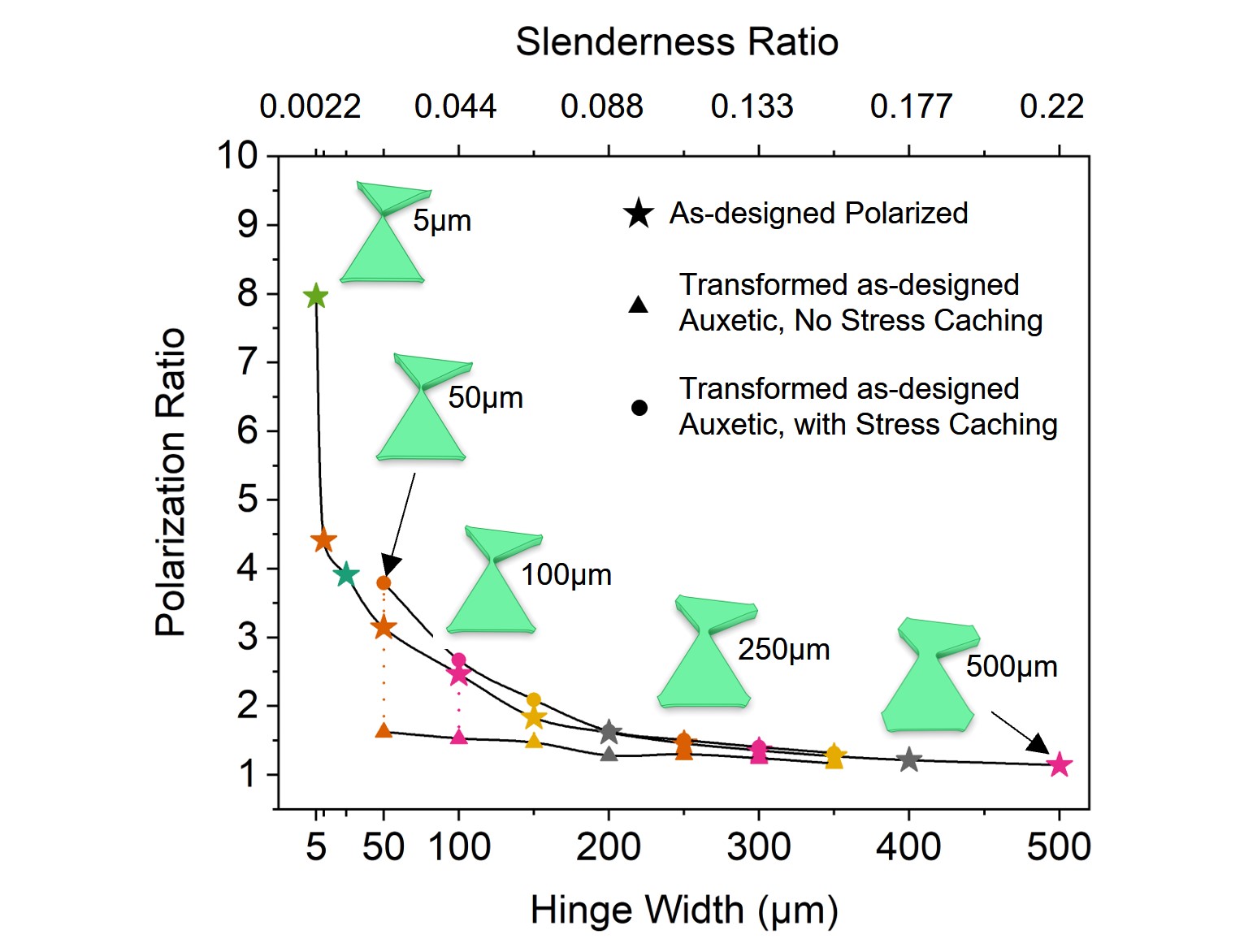}
\caption{{Simulated effects of hinge slenderness on topological polarization in a stiff, elastic-plastic material i.e., utilizing the room temperature mechanical properties of PCLDA-SMP.}}
\label{Figure_S15}
\end{center}
\end{figure}

\section*{Supporting Movies}
\hfill \break
\noindent \textbf{Movie S1.} Finite element simulation results verifying the effectiveness of the proposed kinematic strategy in prescribing a reversible transformation between topologically distinct phases via a set of synchronously applied vector displacements of lattice edge loops.  
\hfill \break

\noindent \textbf{Movie S2.} Finite element simulation results of different-yet-equivalent strategies to determine a vector map that prescribes a reversible kinematic transformation of a TTMM between its auxetic and polarized phases.
\hfill \break

\noindent \textbf{Movie S3}. Experimental demonstration of a complete kinematic cycle of a TTMM lattice via a custom, laser cut Teflon jig that cascades solitary mechanical inputs applied at the L-R sample edges, into a set of synchronously-applied yet individually-prescribed vector displacements of each edge loop. This in turn induces a wholly-determinate biaxial global transformation of the TTMM into a targeted topological phase, courtesy the lattice’s Guest Hutchinson mode.
\hfill \break

\noindent \textbf{Movie S4.} Experimental characterization of the topological edge behavior of a TTMM lattice via quasi-static mechanical indentation of opposite ‘output’ edge pairs i.e., top and bottom edges.
\hfill \break

\noindent \textbf{Movie S5.} Experimental observations of un-cached elastic stresses generated in lattice hinges during a kinematic transformation causing a polarized lattice to ‘snap back’ to its as-fabricated unstressed state. This is only seen in the absence of shape memory and stress caching. Lattices made from a shape memory polymer such as PCLDA-SMP, are stable in any transformed state courtesy their intrinsic ability to cache stresses.